\documentclass[12pt]{iopart}
\usepackage{iopams}
\usepackage{setstack}

\usepackage[english,british]{babel}
\usepackage{graphicx}
\usepackage{hyperref}

\newcommand{\ie}{i.\,e.\ }

\newcommand{\eg}{e.\,g.\ }

\newcommand{\cf}{cf.\ }

\newcommand{\re}{\mathrm{Re}}
\newcommand{\im}{\mathrm{Im}}

\newcommand{\diag}{\mathrm{diag}}

\newcommand{\abs}[1]{|#1|}

\newcommand{\tder}[2]{\case{\rmd^{#2}}{\rmd {#1}^{#2}}}

\begin{document}
\title{Optomechanical deformation and strain in elastic dielectrics}
\author{M. Sonnleitner$^{1,2}$, M. Ritsch-Marte$^{2}$ and H. Ritsch$^{1}$ }
\address{$^{1}$ Institute for Theoretical Physics, University of Innsbruck, Technikerstra\ss e 25, A-6020 Innsbruck, Austria}
\address{$^{2}$ Division for Biomedical Physics, Innsbruck Medical University, M\"{u}llerstra\ss e 44, A-6020 Innsbruck, Austria}
\ead{Matthias.Sonnleitner@uibk.ac.at}
\begin{abstract}
Light forces induced by scattering and absorption in elastic dielectrics lead to local density modulations and deformations. These perturbations in turn modify light propagation in the medium and generate an intricate nonlinear response. We generalise an analytic approach where light propagation in one-dimensional media of inhomogeneous density is modelled as a result of multiple scattering between polarizable slices. Using the Maxwell stress tensor formalism we compute the local optical forces and iteratively approach self-consistent density distributions where the elastic back-action balances gradient- and scattering forces. For an optically trapped dielectric we derive the nonlinear dependence of trap position, stiffness and total deformation on the object's size and field configuration. Generally trapping is enhanced by deformation, which exhibits a periodic change between stretching and compression. This strongly deviates from qualitative expectations based on the change of photon momentum of light crossing the surface of a dielectric. We conclude that optical forces have to be treated as volumetric forces and that a description using the change of photon momentum at the surface of a medium is inappropriate.
\end{abstract}
%
\section{Introduction} \label{sec:_Intro}
\begin{figure}
	\includegraphics[width=8cm]{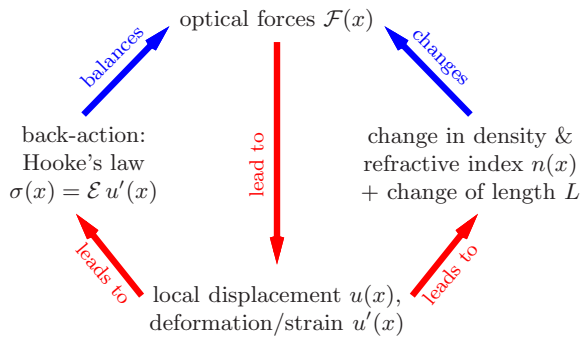}
	\caption{Schematic illustration of the interaction between optical forces and local deformations within elastic media.}
	\label{fig:_inhom_concept}
\end{figure}

As light carries momentum besides energy, its propagation through a polarizable medium is accompanied by forces. Although the momentum of a single light quantum is very small, laser light can generate appreciable forces on the microscopic scale. Optical forces are nowadays routinely used to manipulate and trap particles ranging from single atoms and molecules~\cite{ashkin1970acceleration,hansch1975cooling,phillips1998laser} to plastic beads, biological cells or microbes up to the size of tens of micrometres~\cite{stevenson2010light,padgett2011holographic,thalhammer2011optical}. The mechanical motion of even larger objects such as silica mircodisks or suspended mirrors has been damped and cooled by light forces~\cite{kippenberg2008cavity,wiederhecker2009controlling}. While most of the existing work targets the overall effect on the centre of mass of the particles, it has been shown by us as well as by other groups that these forces do not act homogeneously but exhibit distinct patterns within the medium~\cite{mansuripur2004radiation,zakharian2005radiation,sonnleitner2011optical}. For any elastic medium this leads to local compression or stretching. Of course the modified density also changes the local refractive index and light propagation, which again alters the forces as displayed schematically in figure~\ref{fig:_inhom_concept}. The resulting coupled complex evolution thus obviously requires self-consistent models and solutions~\cite{mansuripur2010resolution}. In addition, as the light mediated interaction is inherently long range, even a small but periodic variation of the refractive index can have a very large overall collective effect coupling distant areas over a large volume.\par
This work is organised as follows: In section~\ref{sec:_inhom_Felder} we first present the basic scattering approach to treat the light propagation in an inhomogeneous refractive medium and use a previously developed formalism based on the free space Maxwell stress tensor to calculate the corresponding local force distribution (section~\ref{sec:_inhom_Kraftdichte}). This method is then used in section~\ref{sec:_Gleichgewicht_Licht_Elast} to develop an iterative scheme to calculate the steady state density and field distribution as a function of geometry and field intensity. In section~\ref{sec:_Numerische_Ergebnisse} we discuss essential physical consequences predicted by our model at the hand of numerical examples. Finally in section~\ref{sec:_photon_momentum} these results are set against common calculations of the total deformation at hand of the change of photon momentum at an interface between to dielectrics.
%
\section{Multiple scattering model of light propagation in inhomogeneous media} \label{sec:_inhom_Felder}
The effective light propagation in a medium can be seen as the result of multiple individual scattering processes, which in general requires intricate numerical treatments, if one cannot make use of material symmetries. Here we restrict ourselves to the simple but still nontrivial case of two incoming counterpropagating plane waves in a transversely homogeneous and linearly polarizable medium. In this limit only forward and backward scattering add up phase coherently, while all amplitudes for transverse scattering average out. From the viewpoint of the forward and backward propagation directions, transverse scattering thus can just be added to an effective absorption rate in the medium. This is certainly not perfectly fulfilled in an actual setup, but still can be expected to give the correct qualitative behaviour, as long as the transverse extensions are much larger than the wavelength of the light. A more realistic treatment, \eg in terms of Gaussian transverse beams, is possible, but greatly complicates the model and would obscure many interesting physical phenomena found in this simple approach.\par
\begin{figure}
	\includegraphics[width=15cm]{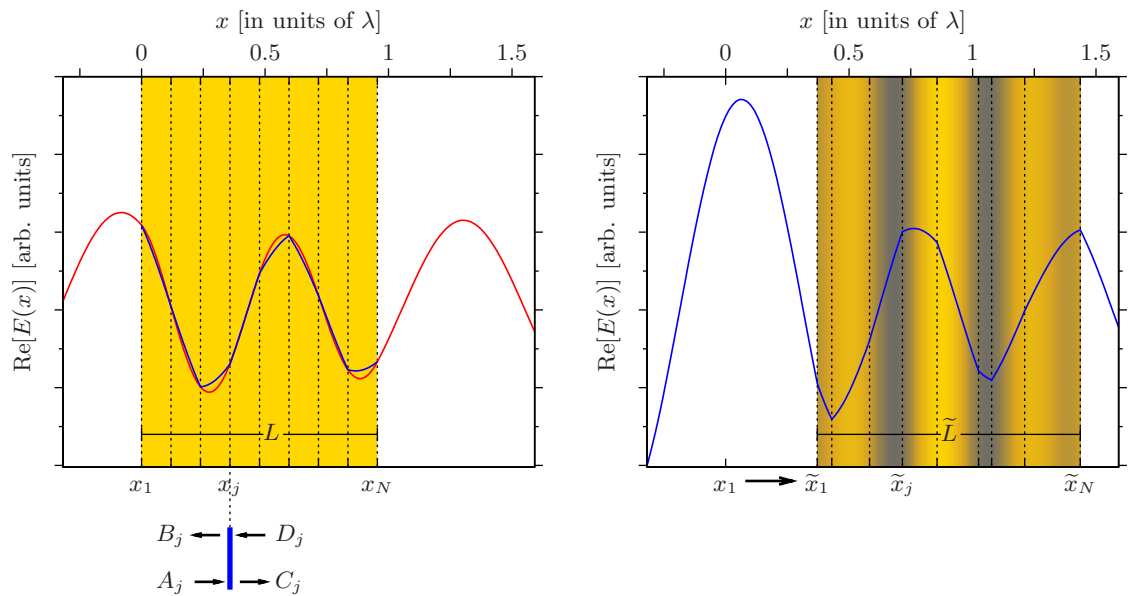}
	\caption{Schematic illustration of the displacement and deformation process $x\mapsto \widetilde{x}=x+u(x)$. The initial medium (left figure) occupies the space $[0,L]$, the slices are marked with dotted lines separated by $d_0=L/(N-1)$. In this case, the field generated by multiple scattering by the beam splitters (blue curve, \cf~\eref{eq:_Intro_ebene_Wellen}) reproduces the solution for a homogeneous medium with refractive index $n$ (red curve), if the coupling is chosen as in~\eref{eq:_zeta}. On the right hand side we see the displaced medium with irregularly spaced slices of the same coupling $\zeta$ and the resulting electric field. The background shading illustrates the change in the distances, \ie the strain $u'(x) = -(\widetilde{\rho}(x)-\rho)/\rho$, with dark colours indicating regions of higher density.}
	\label{fig:_schem_illustration_of_deformation}
\end{figure}

Restricting the dynamics to the forward and backward scattering amplitudes along the propagation directions gives a simple and tractable model for our medium via a one dimensional array of $N$ thin slices at positions $x_1, \ldots, x_N$. Here the spatial behaviour of the electric field $\mathbf{E}(x,t)=\re[E(x) \exp(- \rmi \omega t)]\mathbf{e}_y$ is determined by a 1D Helmholtz equation~\cite{deutsch1995photonic,asboth2008optomechanical, sonnleitner2011optical} 
\begin{equation}\label{eq:_Helmholtz}
	(\partial_x^2+k^2)E(x)=-2 k \zeta E(x) \sum_{j=1}^N \delta(x-x_j).
\end{equation}
The field-induced polarisation density at each slice then is $P(x)=\alpha \eta_\mathrm{A} E(x) \sum_{j=1}^N \delta(x-x_j)$, where we introduced the dimensionless coupling parameter $\zeta = k \eta_\mathrm{A} \alpha/(2 \varepsilon_0)$ proportional to the atomic polarizability $\alpha$ and the areal particle density $\eta_\mathrm{A}$ within the slice. $\varepsilon_0$ is the vacuum permittivity and $k=\omega/c$ the wave number of the optical field. Note that we assumed here that the dipoles in each slice can simply be added up coherently for scattering along the propagation direction. As illustrated in figure~\ref{fig:_schem_illustration_of_deformation}, the equation above is satisfied by interconnected plane wave solutions~\cite{deutsch1995photonic}, here denoted as~
\begin{equation}\label{eq:_Intro_ebene_Wellen}
\fl	E_j(x) := C_j \rme^{\rmi k (x-x_j)} + D_j \rme^{-\rmi k (x-x_j)} = A_{j+1} \rme^{\rmi k (x-x_{j+1})} + B_{j+1} \rme^{-\rmi k (x-x_{j+1})},
\end{equation}
for $x_{j} < x < x_{j+1}$. The amplitudes left and right of a material slice (beam splitter) at position $x_j$ are connected via
\begin{equation}\label{eq:_TrafoMatrix_beamsplitter}
	\pmatrix{ C_j \cr D_j }
	=
	\pmatrix{
		1 + \rmi \zeta 	&   \rmi \zeta 		\cr
		- \rmi \zeta	&	1 - \rmi \zeta	}
	\pmatrix{ A_j \cr B_j }
	 =: \mathrm{M_{BS}} \pmatrix{ A_j \cr B_j }.
\end{equation}
The amplitudes $(A_j,B_j)$ and $(C_{j-1},D_{j-1})$ are coupled by a simple propagation matrix, \ie $(A_j,B_j)^T = \mathrm{P}_{d_j} (C_{j-1},D_{j-1})^T$ with $\mathrm{P}_{d_j}:= \diag\big(\exp(\rmi k d_j),\exp(-\rmi k d_j)\big)$, with the distance $d_j:=x_{j+1}-x_j$, $j=1, \ldots, N-1$.\par
Therefore, the amplitudes to the left of the $(j+1)^\mathrm{th}$ slice are obtained by a simple multiplication of the previous transfer matrices,
\begin{equation}\label{eq:_Amplitudenrelation_BS_allgemein}
	\pmatrix{ A_{j+1} \cr B_{j+1} }
	=
	\mathrm{P}_{d_j} \mathrm{M_{BS}} \cdots \mathrm{P}_{d_2} \mathrm{M_{BS}} \mathrm{P}_{d_1} \mathrm{M_{BS}}
	\pmatrix{ A_{1} \cr B_{1} }.
\end{equation}
The amplitudes $A_1$ and $D_N$ are determined by the amplitudes and phases of waves coming in from the left (\ie $-\infty$) and from the right ($+\infty$), respectively, and constitute boundary conditions on the solutions for the Helmholtz equation~\eref{eq:_Helmholtz}. $B_1$ and $C_N$ are obtained by computing the total reflection and transmission amplitudes via
\begin{equation}\label{eq:_Einfuehrung_rrt}\eqalign{
	\frac{1}{t} \pmatrix{
		t^2 - r_\mathrm{l} r_\mathrm{r}	&	r_\mathrm{r}	\cr
		- r_\mathrm{l}			&	1	}
			=	\mathrm{M_{BS}} \mathrm{P}_{d_{N-1}} \mathrm{M_{BS}} \cdots\mathrm{P}_{d_{2}}  \mathrm{M_{BS}} \mathrm{P}_{d_{1}} \mathrm{M_{BS}}, \\
	B_1=r_\mathrm{l} A_1+t D_N \qquad\mathrm{and}\qquad C_N=t A_1+r_\mathrm{r} D_N.
}\end{equation}
Note that the reflection coefficients for left or right incidence on an inhomogeneous setup usually do not coincide, \ie $r_\mathrm{l}\neq r_\mathrm{r}$, but the transmission amplitude $t$ is independent of the direction of propagation. More details on the properties of these generalised transfer matrices are given in~\ref{sec:_TrafoMatrizen}.\par
For equally spaced, thin polarizable slices we set $x_j = (j-1) d_0$, such that $x_0=0$ and $x_N=(N-1)d_0=:L$ and~\eref{eq:_Amplitudenrelation_BS_allgemein} simplifies to $(A_{j+1}, B_{j+1})^T = \mathrm{T_h}^j (A_1, B_1)^T$, with $\mathrm{T_h}:=\mathrm{P}_{d_0} \mathrm{M_{BS}}$. In an earlier work~\cite{sonnleitner2011optical} we showed that choosing a uniform distance $d_0$ between the slices and setting the coupling parameter
\begin{equation}\label{eq:_zeta}
	\zeta = \frac{\cos(k d_0)-\cos(n k d_0)}{\sin(k d_0)}
\end{equation}
leads to the same optical fields as found inside a medium with refractive index $n$. A sufficiently dense array of beam splitters with spacing $d_0=L/(N-1)$ is then in the limit $N\rightarrow \infty$ indistinguishable from a homogeneous medium of refractive index $n$ and length $L$.\par
A decisive step in this work, which allows us to account for local material density variations, is the introduction of a local \emph{displacement} variable $u(x)$
\begin{equation}~\label{eq:_intro_displacement}
	x_j \mapsto \widetilde{x}_j=x_j+u(x_j),	\qquad j=1, \ldots, N.
\end{equation}
As illustrated schematically in figure~\ref{fig:_schem_illustration_of_deformation}, such shifts alter the local fields as well as the total reflection and transmission properties of the object.\par
The distances between the slices then change as
\begin{equation}
	\widetilde{d}_j-d_0 = \widetilde{x}_{j+1}-\widetilde{x}_j-d_0=u(x_{j+1})-u(x_j)=: \Delta_j.
\end{equation}
A continuous limit can be consistently defined via $u(x_j) \rightarrow u(x)$ for $x\in[0,L]$ to obtain
\begin{equation}\label{eq:_limit_distance_strain}
	\lim_{N\rightarrow \infty} \frac{\Delta_j}{d_0}=\lim_{N\rightarrow \infty} \frac{u(x_j+d_0)-u(x_j)}{d_0} = u'(x).
\end{equation}
In analogy with the theory of elastic deformations, we call $u'$ \emph{strain} or \emph{deformation}~\cite{Landau_Lifschitz_elasticity,Lautrup_continuous_matter}, and the relative change in the initially homogeneous local material density $\rho$ simply reads
\begin{equation}\label{eq:_strain_and_density}
	\frac{\widetilde{\rho}(x)-\rho}{\rho} = - u'(x).
\end{equation}
Let us here comment on the notation we will use for the rest of this work. As defined in the paragraph above~\eref{eq:_zeta}, our coordinates are chosen such that the unperturbed medium occupies the region~$[0,L]$. Introducing a displacement~$u$ then shifts the object to~$[u(0),L+u(L)]$, with $L+u(L)-u(0)=:\widetilde{L}$. But to ease notation, all the quantities such as the electric field strength $E$ or force $F$ shall remain defined with respect to the original position such that \eg $E(0)$ [$E(L)$] always marks the field at the left [right] edge of the medium. The amplitudes at the boundaries then have to be adjusted with corresponding phases, \cf~\eref{eq:_A0_DL_tilde}. This, however, is relevant for mathematical formulations only, physical discussions and figures are unaffected by this detail. In~\eref{eq:_intro_displacement} we introduced a tilde to distinguish the shifted $\widetilde{x}_j$ from the original $x_j$. For most other quantities such as the fields or forces, we will omit this tedious notation. Only the changed length $\widetilde{L}$, the inhomogeneous density $\widetilde{\rho}$~\eref{eq:_strain_and_density} and refractive index $\widetilde{n}$~\eref{eq:_nvonx} still have to be distinguished from their original values $L$, $\rho$ and~$n$.
\par
As mentioned before, defining the coupling $\zeta$ as in~\eref{eq:_zeta} ensures that the solutions of the wave equation~\eref{eq:_Helmholtz} agree with the field inside a homogeneous dielectric at positions $x_j=(j-1)d_0$, if the fields are assumed to agree at $x_1$. In the continuous limit $N\rightarrow \infty$, the latter requirement is always fulfilled~\cite{sonnleitner2011optical}. Interestingly we still preserve this feature for a model with displaced slices, if we choose the following approach:\par
Let, as in~\eref{eq:_Intro_ebene_Wellen}, $E_{j}(x)$ denote the plain wave solution of the Helmholtz equation~\eref{eq:_Helmholtz} and
\begin{equation}
	E_{n_j}(x) = G_j \rme^{\rmi n_j k (x-x_j)} + H_j \rme^{-\rmi n_j k (x-x_j)} \qquad \mathrm{for} \; x_j< x<x_{j+1}
\end{equation}
denote a field defined in the same region, but with a refractive index $n_j$. To obtain the desired equivalence between a stratified dielectric and a set of irregularly spaced slices, we assume for any given $j\in\{1,\ldots,N-2\}$
\begin{equation}\label{eq:_Annahme_BSmed}
	\lim_{x \downarrow x_j} E_{j}(x) = \lim_{x \downarrow x_j} E_{n_j}(x) 
\end{equation}
and demand that with $E(\uparrow\!\! y) \equiv \lim_{x\uparrow y}E(x)$,
\begin{eqnarray}
\fl	& E_{j}(\uparrow\!\! x_{j+1}) = E_{j+1}(\downarrow\!\! x_{j+1}), \qquad & 
			E'_{j}(\uparrow\!\! x_{j+1}) = E'_{j+1}(\downarrow\!\! x_{j+1}) + 2 k \zeta E_{j+1}(\downarrow\!\! x_{j+1}),	\label{eq:_Bedingung_Helmholtz_Loesung} \\
\fl & E_{n_j}(\uparrow\!\! x_{j+1}) = E_{n_{j+1}}(\downarrow\!\! x_{j+1}), \qquad & 
 			 E'_{n_j}(\uparrow\!\! x_{j+1}) = E'_{n_{j+1}}(\downarrow\!\! x_{j+1}), \label{eq:_Bedingung_Fresnel}\\
\fl	& E_{j}(\uparrow\!\! x_{j+1}) = E_{n_j}(\uparrow\!\! x_{j+1}), \qquad &
 		E_{j+1}(\uparrow\!\! x_{j+2}) = E_{n_{j+1}}(\uparrow\!\! x_{j+2}). \label{eq:_Forderung_BSmed}
\end{eqnarray}
The first line shows the conditions that $E_{j}$ and $E_{j+1}$ are solutions of the Helmholtz equation~\eref{eq:_Helmholtz}, \cf \eref{eq:_TrafoMatrix_beamsplitter} or \cite{deutsch1995photonic}, the second line denotes Fresnel's equations for the transition between two dielectrics. In the third line, finally, we demand that the plane wave solutions of~\eref{eq:_Helmholtz} agree with the fields inside the dielectrics at positions $x_{j+1}$ and $x_{j+2}$. This leads to the required, successive coupling between $\zeta$, the distances $d_j = x_j-x_{j-1}$ and $d_{j+1}$, and some indices $n_j$, $n_{j+1}$.\par
Solving~\eref{eq:_Bedingung_Helmholtz_Loesung}-\eref{eq:_Forderung_BSmed} under the assumption~\eref{eq:_Annahme_BSmed} and demanding solutions independent of the field amplitudes results in two conditions, for $j=1, \dots, N-1$
\begin{eqnarray}
	\frac{n_j \sin(k d_j)}{\sin(k n_j d_j)} = \frac{n_{j+1} \sin(k d_{j+1})}{\sin(k n_{j+1} d_{j+1})},	\label{eq:_Bedingung_BSmed_1}\\
	\zeta = \frac{1}{2} \Big[ \frac{\cos(k d_j) - \cos(n_j k d_j)}{\sin(k d_j)} + 
								\frac{\cos(k d_{j+1}) - \cos(n_{j+1} k d_{j+1})}{\sin(k d_{j+1})} \Big]. \label{eq:_Bedingung_BSmed_2}
\end{eqnarray}
One can easily check that these conditions give the known relation~\eref{eq:_zeta} in the equidistant case where $d_j=d_{j+1}\equiv d_0$ and $n_j=n_{j+1}\equiv n$. Unfortunately, we were not able to find solutions with finite values of $d_j\neq d_{j+1}$ for both conditions. Inspired from~\eref{eq:_zeta} one may try 
\begin{equation}\label{eq:_njtest}
	n_j=\case{1}{d_j k}\arccos\big( \cos(d_j k) - \zeta \sin(d_j k) \big).
\end{equation}
to find that this approach satisfies~\eref{eq:_Bedingung_BSmed_2}, but not~\eref{eq:_Bedingung_BSmed_1}. However, choosing $\zeta$ as in~\eref{eq:_zeta}, writing $d_j=d_0(1+\Delta_j/d_0)$, and taking the continuous limit $N\rightarrow \infty$ with $\Delta_j/d_0\rightarrow u'(x)$~\eref{eq:_limit_distance_strain} alters~\eref{eq:_njtest} to
\begin{equation} \label{eq:_nvonx}
	\widetilde{n}(x) = \sqrt{\frac{n^2+u'(x)}{1+u'(x)}},
\end{equation}
satisfying both~\eref{eq:_Bedingung_BSmed_1} and~\eref{eq:_Bedingung_BSmed_2}. With the given inhomogeneous refractive index we can compute the electric field inside a strained, one dimensional dielectric by solving
\begin{equation}\label{eq:_Wellengl_n(x)}
	\big(\partial_x^2+\widetilde{n}^2(x) k^2\big)E(x)=0
\end{equation}
numerically. A comparison with the field computed via the transfer matrix method described in~\eref{eq:_Amplitudenrelation_BS_allgemein} shows excellent agreement, for sufficiently large $N$.\par
Another way to approximate the optical field is to expand the transfer matrices in~\eref{eq:_Amplitudenrelation_BS_allgemein} for small local deformations $\Delta_j$ and then perform the continuous limit. This analytical approximation works sufficiently well for the typically small strain $u'$ obtained in the scope of parameters used in this work. The lengthy results of this approach are presented in~\ref{sec:_ApproxField_Force}, equation~\eref{eq:_Felder_kurz}.\par
Inserting the relation between strain and density modifications~\eref{eq:_strain_and_density} we finally obtain
\begin{equation}\label{eq:_n(x)_and_density}
	\widetilde{n}^2 = \frac{(n^2+1)\rho - \widetilde{\rho}(x)}{2 \rho-\widetilde{\rho}(x)} \simeq n^2 + (n^2-1)\frac{\widetilde{\rho}(x)-\rho}{\rho},
\end{equation}
where we assumed $(\widetilde{\rho}-\rho)/\rho \ll 1$ for the final expansion.
\subsection{Computing the reflection and transmission amplitudes}\label{sec:_rrt}
To find solutions for the fields inside the medium with refractive index distribution $\widetilde{n}(x)$, one needs to specify initial values. As discussed for the discrete system in~\eref{eq:_Einfuehrung_rrt}, the medium can be described in terms of a transfer matrix such that $B_0=r_\mathrm{l} A_0+t D_0$ and $C_L=t A_0+r_\mathrm{r} D_L$, if the electric fields outside the medium are given as $E(x)=A_0 \exp(\rmi k x) + B_0 \exp(-\rmi k x)$ for $x\leq 0$ and $E(x)=C_L \exp(\rmi k (x-L)) + D_L \exp(-\rmi k (x-L))$ for $x\geq L$. The amplitudes $A_0$ and $D_L$ are determined by the intensities $I_{l,r}$ and phases $\phi_{l,r}$ of the fields incident from the left and right and the displacement $u$, as
\begin{equation}\label{eq:_A0_DL_tilde}
	A_0 = \sqrt{\case{2 I_\mathrm{l}}{\varepsilon_0 c}} \rme^{\rmi \phi_\mathrm{l}} \rme^{\rmi k u(0)}
	\qquad\mathrm{and}\qquad
	D_L = \sqrt{\case{2 I_\mathrm{r}}{\varepsilon_0 c}} \rme^{\rmi \phi_\mathrm{r}} \rme^{-\rmi k (L+u(L))} .
\end{equation}
Therefore, the initial conditions for solutions of~\eref{eq:_Wellengl_n(x)} are $E(0)=A_0+B_0$ and $E'(0)=\rmi k(A_0-B_0)$. \par
But obviously, the reflection and transmission coefficients $r_\mathrm{l}$, $r_\mathrm{r}$ and $t$ strongly depend on the refractive index $\widetilde{n}(x)$. To calculate those one can either use some approximations, \cf~\ref{sec:_ApproxField_Force}, equation~\eref{eq:_r_und_t_inhomogen}, or solve the field equation~\eref{eq:_Wellengl_n(x)} for specially chosen boundary values, \eg 
\begin{equation}\eqalign{
\fl	E^{[1]}(0)= \rmi E^{[1]\prime}(0)/k = t D_L \quad\Rightarrow\quad E^{[1]}(L)=D_L(r_\mathrm{r}+1), \, E^{[1]\prime}(L)=\rmi k D_L(r_\mathrm{r}-1), \\
\fl	E^{[2]}(L)= -\rmi E^{[2]\prime}(L)/k = t A_0 \quad\Rightarrow\quad E^{[2]}(0)=A_0(1+r_\mathrm{l}), \, E^{[2]\prime}(0)=\rmi k A_0(1+r_\mathrm{l}),
}\end{equation}
allowing the easy computation of $r_\mathrm{l}$, $r_\mathrm{r}$ and $t$.\par
It is easy to see that if $\widetilde{n}(x)$ is symmetric, \ie $\widetilde{n}(x)=\widetilde{n}(L-x)$, $x\in[0,L]$, then a beam entering from the left experiences the same medium as one from the right and hence $r_\mathrm{l}=r_\mathrm{r}$. Note that for the homogeneous case where $u'=0$ and $\widetilde{n}=n$, we recover the usual~\cite{Born_priciples_of_optics}
\begin{equation}\label{eq:_r_und_t_homogen}\eqalign{
	t_\mathrm{h} =\frac{2 n}{2 n \cos(n k L)-\rmi (n^2+1) \sin(n k L)},	\\
	r_\mathrm{h} =\frac{\rmi (n^2-1) \sin(n k L)}{2 n \cos(n k L)-\rmi (n^2+1) \sin(n k L)}.
}\end{equation}
%
\section{Light forces in an inhomogeneous medium}\label{sec:_inhom_Kraftdichte}
In general, the total electromagnetic force on an object embedded in vacuum is given by~\cite{Jackson_class_electrodyn}
\begin{equation}
	\mathrm{F}_\alpha=\oint_\mathcal{A} \sum_\beta \mathrm{T}_{\alpha \beta}\, \mathrm{n}_\beta \rmd A,
\end{equation}
where $\mathcal{A}$ denotes the surface of the object, $\mathrm{n}$ is the normal to $\mathcal{A}$ and $\mathrm{T}_{\alpha \beta}$ is the Maxwell stress tensor
\begin{equation}
	\mathrm{T}_{\alpha \beta} = \varepsilon_0 \mathrm{E}_\alpha \mathrm{E}_\beta + \case{1}{\mu_0} \mathrm{B}_\alpha \mathrm{B}_\beta - \case{1}{2} \delta_{\alpha,\beta} \big(\varepsilon_0 \mathrm{E}^2+ \case{1}{\mu_0} \mathrm{B}^2\big).
\end{equation}
Using two planes orthogonal to the direction of propagation (\ie the $x$-axis) as integration surfaces and the plane wave fields defined in~\eref{eq:_Intro_ebene_Wellen}, the time-averaged optical force per area (pressure) on the $j^\mathrm{th}$ slice simply reads~\cite{xuereb2009scattering}
\begin{equation}\label{eq:_Kraft_Fj}
	F_j=\frac{\varepsilon_0}{2} \Big( \abs{A_j}^2 + \abs{B_j}^2 - \abs{C_j}^2 - \abs{D_j}^2 \Big). 
\end{equation}
Following the beam splitter relation in~\eref{eq:_TrafoMatrix_beamsplitter} we rewrite $C_j = (1+\rmi \zeta) A_j + \rmi \zeta B_j$ and $D_j=-\rmi \zeta A_j + (1-\rmi \zeta) B_j$ to obtain
\begin{equation}\label{eq:_Kraft_Fj_anders}
	F_j=-\varepsilon_0 \Big(\abs{\zeta (A_j+B_j)}^2-\im[\zeta(A_j+B_j)(A_j-B_j)^\ast]\Big).
\end{equation}
Taking the naive limit $\lim_{N\rightarrow \infty} F_j$ would give a vanishing force per slice as $\lim_{N\rightarrow \infty} \zeta = 0$, \cf~\eref{eq:_zeta} with $d_0=L/(N-1)$. But assigning each slice to one $N^\mathrm{th}$ of the object's total length $L$ we can define a force density $\mathcal{F}(x):= \lim_{N\rightarrow \infty} N F_j/L$ and use
\begin{equation}\label{eq:_zeta_limes}
	\lim_{N \rightarrow \infty}{\case{N}{L}\zeta}= k\case{n^2-1}{2}.
\end{equation}
Following the derivation of the inhomogeneous refractive index~\eref{eq:_nvonx} we can replace $A_j+B_j=E_j(\uparrow\!\! x_{j+1}) = E_{n_j}(\uparrow\!\! x_{j+1}) \rightarrow E(x)$~\eref{eq:_Forderung_BSmed} and in the continuous limit it is reasonable to set $A_j-B_j = -\rmi E_j'(\uparrow\!\! x_{j+1})/k \rightarrow -\rmi E'(x)/k$, if $E(x)$ is a solution of the wave equation~\eref{eq:_Wellengl_n(x)}. Hence we obtain the local optical force density
\begin{equation}\label{eq:_Kraftdichte_n(x)}
	\mathcal{F}(x) = \case{\varepsilon_0}{2} \re\big[(n^2-1) E(x) \big(E'(x)\big)^\ast \big],
\end{equation}
where we used the algebraic limit theorem for $\lim_{N\rightarrow \infty} N \abs{\zeta}^2/L=0$. Again, the results from the formula above are fully consistent with forces computed using a large but finite number of slices~\eref{eq:_Amplitudenrelation_BS_allgemein} and~\eref{eq:_Kraft_Fj} as well as an analytical approximation presented in~\ref{sec:_ApproxField_Force}, equation~\eref{eq:_Kraftdichte_Einzelterme}.
\par
\begin{figure}
	\includegraphics[width=12cm]{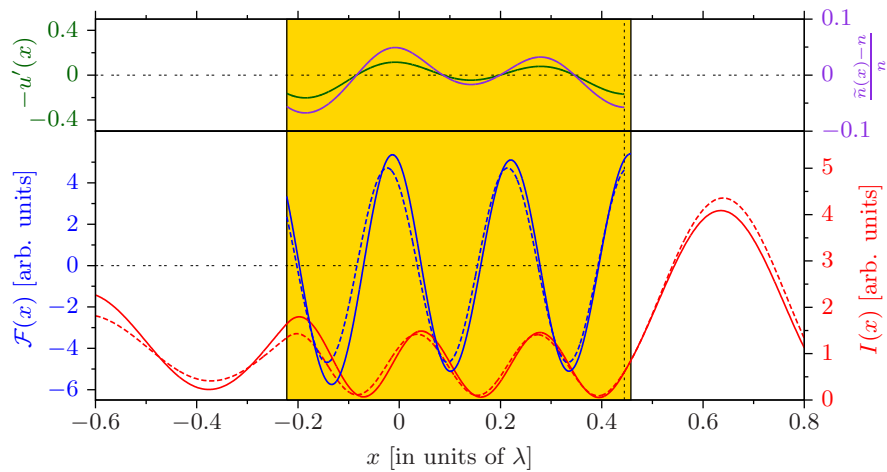}
	\caption{Illustration of the influence of a given strain $u'$ (green) on the optical force density (blue) and the local intensity (red). The dotted lines show the homogeneous case $u'=0$, the continuous lines represent the perturbations computed using the refractive index distribution $\widetilde{n}(x)$ (purple). The reflection and transmission amplitudes here change from $r_\mathrm{h}\simeq-0.29-0.29\rmi$, $t_\mathrm{h}\simeq-0.64+0.59\rmi$ to $r_\mathrm{l}\simeq-0.20-0.21\rmi$, $r_\mathrm{r}\simeq-0.15-0.24\rmi$, $t\simeq-0.76+0.51\rmi$. Please note that in this case the strain is not chosen such that it balances the optical forces as in~\eref{eq:_Gleichgewicht_Licht_Elast}.}
	\label{fig:_hom_vs_inhom}
\end{figure}
In figure~\ref{fig:_hom_vs_inhom} we compare the intensity and optical forces in a medium with homogeneous refractive index $n$ to fields and forces obtained by solving~\eref{eq:_Wellengl_n(x)} and~\eref{eq:_Kraftdichte_n(x)}, respectively, for a given strain $u'$.
\subsection{Identification of radiation pressure and dipole force components inside a homogeneous dielectric}
In the case of a medium with uniform refractive index $n$, \ie with $u'\equiv0$, the force density computed from~\eref{eq:_Kraftdichte_n(x)} can be identified with established expressions for optical forces on dielectric test particles. There, the time-averaged force on a dipole at position $x_0$ in the external field $E(x,t)=\re\{ E(x)\exp(-\rmi\omega t)\}$ with $E(x)=\abs{E(x)}\exp(-\rmi \varphi(x))$ reads~\cite{Cohen-Tannoudji_atomic_motion}
\begin{equation}\label{eq:_Kraft_CohTan}
	F_\mathrm{L} = \case{1}{4} \partial_x \abs{E(x_0)}^2 \re \alpha - \case{1}{2} \abs{E(x_0)}^2 \partial_x\varphi(x_0) \im \alpha \,,
\end{equation}
where $\alpha$ is the polarizability of the dipole. The first term proportional to $\re \alpha$ is often referred to as \emph{dipole} or \emph{gradient force}, the (dissipative) term proportional to $\im \alpha$ is called \emph{radiation pressure} or \emph{scattering force}~\cite{Cohen-Tannoudji_atomic_motion}.\par
Inside a homogeneous dielectric we may write the spatial component of the electric field as $E(x) = G \exp(\rmi n k (x-x_0)) + H \exp(-\rmi n k (x-x_0))$, where the amplitudes $G$ and $H$ are chosen such that Fresnel's conditions at the boundary of the object are met. Using this field and rewriting~\eref{eq:_Kraft_CohTan} into a force density on particles with volume density $\eta_\mathrm{V}$ located at $x_0$ leads to
\begin{equation}\label{eq:_Kraft_CohTan_x0}
	\mathcal{F}_\mathrm{L}(x_0) = - \eta_\mathrm{V} k \im\{G H^\ast\} \re\{\alpha n^\ast\} + \case{k}{2} \eta_\mathrm{V} \big(\abs{G}^2 - \abs{H}^2\big) \im\{\alpha n^\ast\}.
\end{equation}
As the coupling parameter in~\eref{eq:_Helmholtz} is defined as $\zeta = k \eta_\mathrm{A} \alpha/(2 \varepsilon_0)$, with $\eta_\mathrm{A}$ denoting the areal particle density in each of the $N$ slices, we may write using~\eref{eq:_zeta} and~\eref{eq:_zeta_limes}
\begin{equation}\label{eq:_alpha_zeta_CohTan}
	\eta_\mathrm{V} \alpha = \lim_{N \rightarrow \infty} \case{N}{L} \eta_\mathrm{A} \alpha = \lim_{N \rightarrow \infty} 2 \case{N}{k L} \varepsilon_0 \zeta =\varepsilon_0 (n^2-1).
\end{equation}
This relation also resembles the Lorentz-Lorenz relation for the case of a thin gas~\cite{Born_priciples_of_optics}, where the individual dipoles do not directly interact with each other.\par
It can easily be checked that the force density in~\eref{eq:_Kraft_CohTan_x0} together with the Lorentz-Lorenz relation~\eref{eq:_alpha_zeta_CohTan} gives exactly the same result as the force computed from~\eref{eq:_Kraftdichte_n(x)}, if we insert the same field for a homogeneous dielectric. This demonstrates that our approach to calculate fields and forces from multiple scatterers is consistent with well known results derived from more general assumptions.
\subsection{Integrated force and trap formation}\label{sec:_Fallenposition}
To compute the total force on an extended dielectric in a standing wave we can use the same derivation as for the force on an infinitesimal slice in~\eref{eq:_Kraft_Fj} and get
\begin{equation}\label{eq:_Kraft_Med_1}
	F_\mathrm{tot}=\frac{\varepsilon_0}{2} \Big( \abs{A_0}^2 + \abs{r_\mathrm{l} A_0+t D_L}^2 - \abs{r_\mathrm{r} D_L+t A_0}^2 - \abs{D_L}^2 \Big),
 \end{equation}
with $A_0$ and $D_L$ denoting the amplitudes at the object's left and right boundaries, as given in~\eref{eq:_A0_DL_tilde}. Defining the position of the centre of the object $x_0=(u(0)+L+u(L))/2$, we can express the total force in terms of $\xi:=x_0 + (\phi_\mathrm{l} - \phi_\mathrm{r})/(2 k)$
\begin{equation}\label{eq:_Kraft_Med_2}
	F_\mathrm{tot}(\xi)=\case{1}{c} \big(I_\mathrm{l} s_\mathrm{l} - I_\mathrm{r} s_\mathrm{r} + 2 v \sqrt{I_\mathrm{l} I_\mathrm{r}} \cos(2 k \xi + \psi) \big),
\end{equation}
with $s_\mathrm{l} = 1+\abs{r_\mathrm{l}}^2-\abs{t}^2$, $s_\mathrm{r} = 1+\abs{r_\mathrm{r}}^2-\abs{t}^2$, $v = \abs{r_\mathrm{l} t^\ast - r_\mathrm{r}^\ast t}$, $\psi = \arg(r_\mathrm{l} t^\ast - r_\mathrm{r}^\ast t)$.\par
If $4 I_\mathrm{l} I_\mathrm{r} v^2 \geq (I_\mathrm{l} s_\mathrm{l} - I_\mathrm{r} s_\mathrm{r})^2$, the force vanishes at every position $\xi_0 \in \Xi_{+} \cup \Xi_{-}$ where
\begin{equation}\label{def:_Fallenpositionen}\eqalign{
	\Xi_{+} := \Big\{ \case{m \pi}{k} + \widehat{\xi}_0, m \in \mathbb{Z} \Big\} 
\qquad\mathrm{and}\qquad
	\Xi_{-} := \Big\{ \case{m \pi - \psi}{k} - \widehat{\xi}_0, m \in \mathbb{Z}\Big\}, \\
\widehat{\xi}_0 : = \frac{1}{2 k} \Big( \arccos\Big[ \frac{I_\mathrm{r} s_\mathrm{r}-I_\mathrm{l} s_\mathrm{l}}{2 v \sqrt{I_\mathrm{l} I_\mathrm{r}}}\Big]-\psi\Big).
}\end{equation}
Using some trigonometric properties one can show that stable trapping positions are those defined in the set $\Xi_{+}$. For $\xi_0 \in \Xi_-$ we find $F_\mathrm{tot}^\prime(\xi_0)>0$ and hence $\Xi_-$ is a set of unstable trapping position. Linearising the total force in~\eref{eq:_Kraft_Med_2} around stable trapping positions $\xi_0 \in \Xi_+$ leads to a trap stiffness of
\begin{equation}\label{eq:_trap_stiffness}
	\kappa=\case{2 k}{c}\sqrt{4 I_\mathrm{l} I_\mathrm{r} v^2 - (I_\mathrm{l} s_\mathrm{l} - I_\mathrm{r} s_\mathrm{r})^2}.
\end{equation}
Since the reflection and transmission coefficients strongly depend on the object's size, one finds that the parameter $\psi$ can change its sign abruptly for certain values of $L$. This leads to sudden jumps between a low- and high-field-seeking behaviour for a trapped object~\cite{zemanek2003theoretical,vcivzmar2006optical,stilgoe2008effect,sonnleitner2011optical}. Figure~\ref{fig:_trap_stiffness} shows how trap position and trap stiffness is changed by the strain induced on the object by optical forces.
%
\section{Self-consistent balancing of optical force and elastic back-action}~\label{sec:_Gleichgewicht_Licht_Elast}
In the previous chapters~\ref{sec:_inhom_Felder} and~\ref{sec:_inhom_Kraftdichte} we found expressions for the local fields and the local optical force densities in deformed, dielectric media. But depending on the given elastic properties, the strain will result in stress which typically tries to compensate the external volumetric forces.\par 
In this chapter we will investigate the behaviour of a linear elastic, dielectric object subjected to the optical forces described by~\eref{eq:_Kraftdichte_n(x)}. More precisely, we will provide a framework to compute the equilibrium configuration between the optical forces and the elastic counter reaction in a self-consistent manner. In our computations we will assume only optical forces and neglect thermal or piezoelectric effects as well as surface tension.\par
Mechanical equilibrium between some general volume force density $f$ and the resulting stress denoted by the tensor $\sigma$ is given by Cauchy's equilibrium equation~\cite{Lautrup_continuous_matter}
\begin{equation}\label{eq:_Chauchy_equilibrium}
	f_i+\sum_j \partial_j \sigma_{ij} = 0,
\end{equation}
for $i,j$ denoting the coordinates of the system. Since the model discussed here considers only one relevant dimension, this equilibrium equation simplifies to $f+\partial_x\sigma=0$. The constitutive relation for a linear elastic, one dimensional object simply reads $\sigma = \mathcal{E} u'$, where $\mathcal{E}$ is Young's modulus and $u'$ is the local strain~\cite{Lautrup_continuous_matter}.\par
Hence we see that an equilibrium between the optical force density and the elastic strain requires
\begin{equation}\label{eq:_Gleichgewicht_Licht_Elast}
	\mathcal{F}(x) + \mathcal{E} u''(x) = 0,
\end{equation}
at every position $x\in [0,L]$, with $\mathcal{F}$ being a solution of~\eref{eq:_Kraftdichte_n(x)}. Note that the electric field computed from~\eref{eq:_Wellengl_n(x)} also depends on the amplitudes at the edges of the object and therefore also on the displacement $u$, \cf~\eref{eq:_A0_DL_tilde}.\par
Solving~\eref{eq:_Gleichgewicht_Licht_Elast} for an equilibrium requires boundary conditions on the displacement $u$ and the strain $u'$ which are determined by the given setup. The displacement is fixed by the assumed trapping mechanism, \eg if the object is trapped by a standing wave, we have to fulfil $(u(0)+L+u(L))/2 \in \Xi_{+}$. But note that $\widehat{\xi}_0$ depends on the reflection and transmission coefficients and hence also on the deformation $u'$, \cf~\eref{def:_Fallenpositionen}.\par
The strain has to be chosen such that the stress $\sigma = \mathcal{E} u'$ at each surface balances external \emph{surface} forces~\cite{Lautrup_continuous_matter}. Assuming for the moment an object subjected to volumetric optical forces only, we integrate the equilibrium equation~\eref{eq:_Gleichgewicht_Licht_Elast} at obtain
\begin{equation}\label{eq:_Gleichgewicht_Licht_Elast_integriert}
	0=\int_0^L \big(\mathcal{F}(x)+\sigma'(x)\big)\rmd x = F_\mathrm{tot} + \sigma(L)-\sigma(0).
\end{equation}
For an object trapped by light fields, we get $F_\mathrm{tot}=0$ and $\sigma(0)=\sigma(L)=0$, due to the lack of surface pressure. In chapter~\ref{sec:_RB1}, however, we fix the slab by an external mechanism balancing the total optical force via surface interaction. Hence if the left boundary of the slab is retained at $x=0$ (\ie $u(0)=0$), then $\sigma(L)=0$ and $\sigma(0)=F_\mathrm{tot}$.\par
To solve equation~\eref{eq:_Gleichgewicht_Licht_Elast} numerically, we use an iterative approach where the equilibrium condition is rewritten in the form
\begin{equation}\label{eq:_Gleichgewicht_Licht_Elast_iterativ}
	\mathcal{F}(x)[u_i,u_i'] + \mathcal{E} u_{i+1}''(x) = 0,
\end{equation}
with $u_i$ and $u_i'$ denoting the displacement and strain obtained by the $i^\mathrm{th}$ iteration step and $\mathcal{F}(x)[u_i,u_i']$ is the force density computed using $u_i$ and $u_i'$. The updated $u_{i+1}'$ and $u_{i+1}$ can then be obtained by simple numerical integration, with integration constants chosen in accordance to the boundary conditions of the used setup. With the updated optical force densities $\mathcal{F}(x)[u_{i+1},u_{i+1}']$ one can compute the next step of the iteration. An obvious choice for initial values is a homogeneously shifted distribution (\ie a constant $u_0$) with a given starting length $L$ and refractive index $n$.\par
This iterative scheme proved to be sufficiently exact but significantly faster compared to other methods of solving nonlinear equations, such as Newton's method. We also confirmed our computations with force densities obtained from the transfer matrix approach in~\eref{eq:_Amplitudenrelation_BS_allgemein} and the analytic approximation described in~\ref{sec:_ApproxField_Force}, respectively.
%
\section{Examples and physical interpretation}\label{sec:_Numerische_Ergebnisse}
In our basic considerations above we always assumed the object to be exposed to two counterpropagating laser beams of the same, linear polarisation forming a standing wave. These results can easily be extended to describe situations with only one incident beam or with two counterpropagating beams of different polarisation. In the latter case, one has to calculate the intensities and forces separately for each polarisation direction.\par
In this section we will present four showcase examples to give insight into the large variety of possible results. The first two examples deal with the case where an object is trapped by two counterpropagating beams. The latter examples will treat the case where the object is illuminated by only one beam and externally fixed at one end. For each of the given examples one has to specify the boundary conditions for $u$ and $u'$, as discussed in chapter~\ref{sec:_Gleichgewicht_Licht_Elast}.\par
For all considered setups we will see that the interaction between optical forces and elastic back-action strongly depends on the ratio between the initial length $L$ and the wavelength of the deformation beam in the unperturbed medium, $\lambda/\re[n]$. Concerning the intensity and the elastic properties of the medium we find that all results grow linearly in $(I_\mathrm{l}+I_\mathrm{r})/(\mathcal{E} c)$, at least in the scope of parameters where a solution of the equilibrium equation~\eref{eq:_Gleichgewicht_Licht_Elast} could be obtained within a reasonable error tolerance. That is why the local intensities, $I(x)=\varepsilon_0 c \abs{E(x)}^2/2$, and force densities are given in units proportional to $\mathcal{E}$ in the upcoming figures. Note that the numbers used in the simulations are unrealistic in order to exaggerate the effects, since an intensity of $I=0.1 \mathcal{E}c$ would imply $I\simeq 30 \mathrm{W}/\mu\mathrm{m}^2$ for $\mathcal{E}\sim 1\mathrm{MPa}$.
\subsection{Example: Object trapped by two counterpropagating beams}\label{sec:_RB0}
\begin{figure}
	\includegraphics[width=15cm]{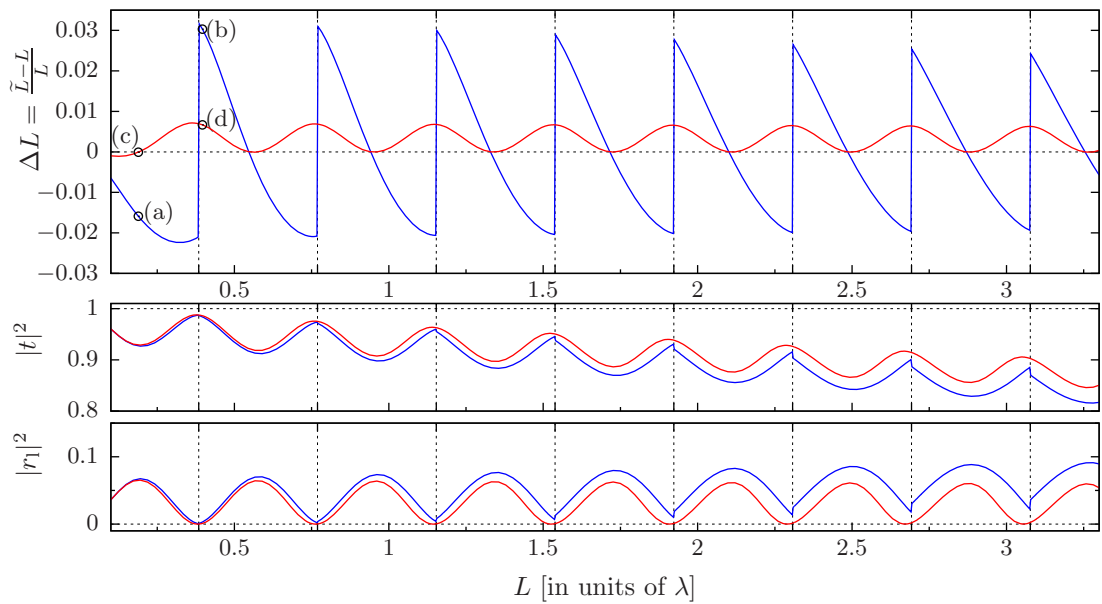}
	\caption{Relative length change, transmission $\abs{t}^2$ and reflection $\abs{r_\mathrm{l}}^2$ for a dielectric slab trapped by two plane waves forming a standing wave (blue) or having orthogonal directions of polarisation (red lines). The incoming intensities and Young's modulus are related as $I_\mathrm{l}=I_\mathrm{r}=0.05 \mathcal{E} c$, where $c$ is the speed of light in vacuum; the refractive index is chosen as $n=1.3+0.0025 \rmi$. The discontinuities in the standing-wave case stem from abrupt jumps in the stable trapping positions at $L= m \lambda/(2 \re[n])$, $m\in \mathbb{N}$ (grid lines). The circles indicate the values used for the examples in figure~\ref{fig:_Bsp_RB0_Int1} and~\ref{fig:_Bsp_RB0_Int0}.}
	\label{fig:_dL_RB0}
\end{figure}
As argued above, an object with initial length $L$ subjected to optical forces will in general experience local deformations and an overall length change. Figure~\ref{fig:_dL_RB0} shows the relative length change $\Delta L=(\widetilde{L}-L)/L$ for different initial lengths $L$ in the cases where an object is trapped in a standing wave (blue lines) or by two beams of orthogonal polarisation (red curves) and equal intensity. Obviously, both configurations are symmetric regarding an inversion of $x$ at the centre of the object and therefore we find $r_\mathrm{l}=r_\mathrm{r}$, as mentioned in section~\ref{sec:_rrt}.\par
Surprisingly, in the standing wave case we observe abrupt switching from strong compressive to  stretching behaviour around certain initial lengths. A comparison with chapter~\ref{sec:_Fallenposition} and earlier discussions in~\cite{sonnleitner2011optical} shows that these switches are concurrent with jumps of the stable trapping position $\xi_0$. Generally speaking, objects with small values of $k L$ are trapped at local maxima of the intensity in the standing wave. But for larger objects, the term $\psi$ in~\eref{def:_Fallenpositionen} abruptly changes its sign and the object seeks positions centred around intensity minima. As indicated by the dotted grid lines, these jumps occur at lengths of minimal reflection $\abs{r_\mathrm{h}}^2$ and maximal transmission $\abs{t_\mathrm{h}}^2$~\eref{eq:_r_und_t_homogen}, \ie at $L = m\lambda/(2 \re[n])$, $m\in \mathbb{N}$.\par
For larger objects we observe a slight decay of the maximal relative length change, which is found also for computations with $\im[n]=0$ and thus is not caused by additional radiation pressure only. However, a glance at the reflectivity and transmission shows that the self consistent deformation prevents configurations with zero reflectivity which would usually result in maximum elongation or compression. For the given standing wave trap we generally observe that deformations computed with the steady state equation~\eref{eq:_Gleichgewicht_Licht_Elast} tend to increase reflectivity and decrease transitivity compared to a homogeneous medium, for both $\im[n]=0$ or $\im[n]\neq 0$.\par
\begin{figure}
	\includegraphics[width=9cm]{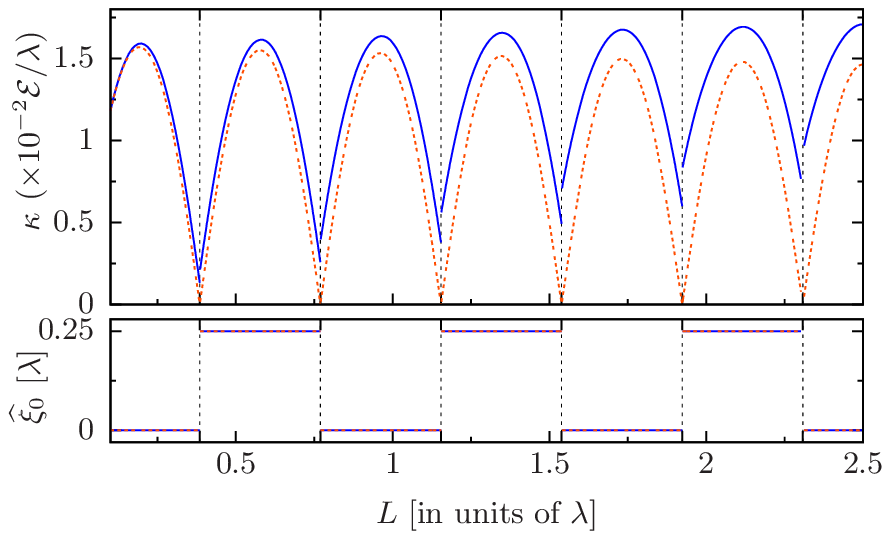}
	\caption{Trap stiffness $\kappa$~\eref{eq:_trap_stiffness} and stable trapping position $\widehat{\xi}_0\in\Xi_+$~\eref{def:_Fallenpositionen} for an object trapped in a standing wave. The same parameters are used as in fig.~\ref{fig:_dL_RB0}, \ie $I_\mathrm{l}=I_\mathrm{r}=0.05 \mathcal{E} c$ and $n=1.3+0.0025 \rmi$. A comparison of the blue curves denoting equilibrium solutions of~\eref{eq:_Gleichgewicht_Licht_Elast} with the orange, dashed lines computed for the homogeneous objects shows that the deformation induced by optical forces significantly increases the total trapping strength. For the given case where $I_\mathrm{l}=I_\mathrm{r}$, the trap positions remain unchanged.}
	\label{fig:_trap_stiffness}
\end{figure}
In figure~\ref{fig:_trap_stiffness} we compare the trap stiffness $\kappa$ from~\eref{eq:_trap_stiffness} and the trapping position $\widehat{\xi}_0$ from~\eref{def:_Fallenpositionen} between unperturbed and self-consistently deformed objects. We find that the deformation significantly increases the trap stiffness, even if the total size of the object remains unchanged, \cf figure~\ref{fig:_dL_RB0}. An elastic object in a standing wave therefore assists in enforcing its own trap, just like a rabbit who starts to dig when captured in a pit.  Note that for a compressible slab in a standing wave there is no critical length of zero trap stiffness, but deformation always leads to a stable trapping position.\par
\begin{figure}
	\includegraphics[width=15cm]{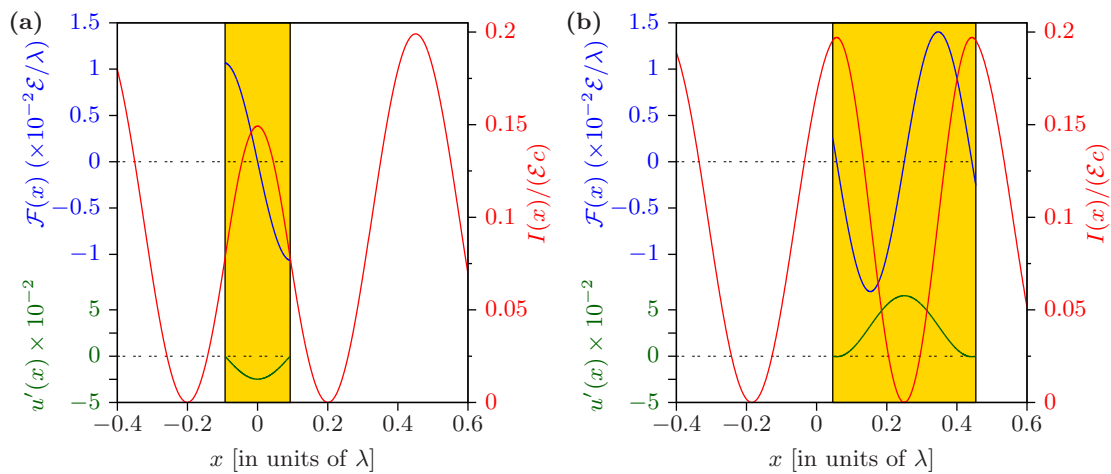}
	\caption{Dielectric objects trapped in a standing wave configuration with $I_\mathrm{l}=I_\mathrm{r}=0.05 \mathcal{E} c$. The initial lengths are chosen as $L\simeq 0.19 \lambda$~(a) and $L\simeq 0.4 \lambda$~(b); $n=1.3+0.0025  \rmi$. As mentioned in~\eref{eq:_strain_and_density}, positive strain $u'(x)$ (green curves) denotes larger distances between the particles and hence a reduction of material density. The density is increased at positions where the force density (blue lines) $\mathcal{F}$ is zero and $\mathcal{F}'<0$, \ie where the light intensity (depicted red) has a maximum value. As indicated by the marks in figure~\ref{fig:_dL_RB0} and verifiable from the signs of $u'(x)$, the object on the left gets squeezed whereas the right hand side example shows a stretched object.}
	\label{fig:_Bsp_RB0_Int1}
\end{figure}
Two examples for optical force densities and the associated deformation in the standing wave setup are shown in figure~\ref{fig:_Bsp_RB0_Int1}. There we see that the strain is negative (\ie the material density is increased) at positions where the force density changes from positive values, denoting a force pushing to the right, to negative values associated with a force pushing to the left. One can clearly see the difference between the compressive situation~(a) where the object is trapped at maximal intensity and situation~(b), where the trapping occurs at minimal intensity and the object experiences a stretching force.\par
\begin{figure}
	\includegraphics[width=15cm]{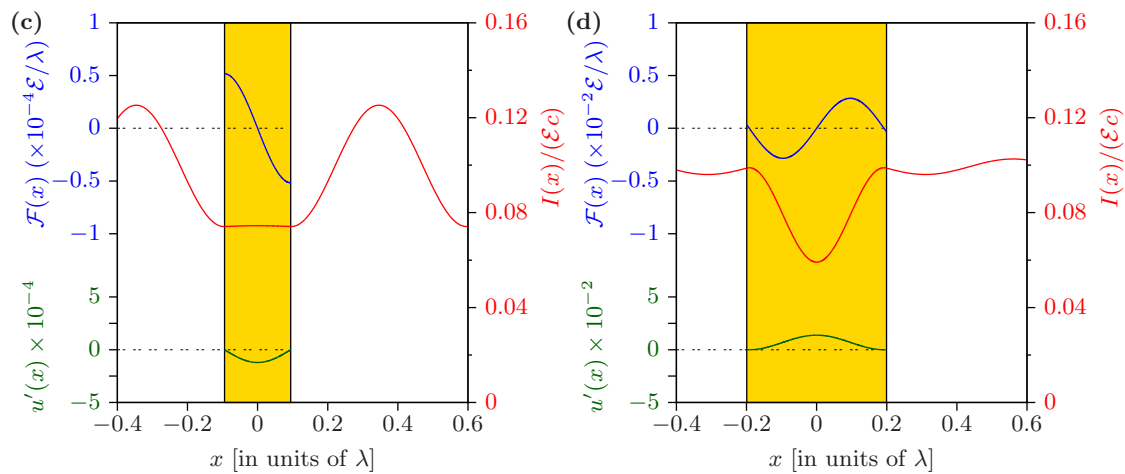}
	\caption{Dielectric objects trapped by two counterpropagating beams with orthogonal polarisation and $I_\mathrm{l}=I_\mathrm{r}=0.05 \mathcal{E} c$. As in figure~\ref{fig:_Bsp_RB0_Int1}, the initial lengths are $L\simeq 0.19 \lambda$~(a) and $L\simeq 0.4 \lambda$~(b); $n=1.3+0.0025 \rmi$. As can be seen from figure~\ref{fig:_dL_RB0}, the length of the first example is chosen such that the intensities (red curves) inside the medium almost add up to a constant value. Hence the force densities (blue curves) and the self-consistent deformation (green lines) practically vanish (please note the changed scale on the axes). For the second example we obtain a stretching of $\Delta L \simeq 6.6 \times 10^{-3}$.}
	\label{fig:_Bsp_RB0_Int0}
\end{figure}

To trap a dielectric slab with two non-interfering plane waves of orthogonal polarisation, the intensities of said beams have to be equal, \ie $I_\mathrm{l}=I_\mathrm{r}$. In this case there are certain starting lengths $L\simeq(2 m+1)/(4 \re[n])$, $m\in \mathbb{N}_0$, for which the intensities inside the unperturbed object add up to a constant value. Hence for these specific lengths the dominant gradient forces add up to zero and the object's length remains unchanged. Apart from that we find an expanding behaviour for $L>\lambda/(4 \re[n])$, as can also be seen in the examples in figure~\ref{fig:_Bsp_RB0_Int0}.\par
In the case of non-interfering beams, figure~\ref{fig:_dL_RB0} also shows that the reflectivity and transitivity is no longer periodic in the object's length $L$. For the reflectivity we find that the zeros are shifted from $L=m\lambda/(2 \re[n])$, $m\in \mathbb{N}$, towards smaller lengths and no longer coincide with lengths of maximal elongation.
\subsection{Example: Object fixed at the left boundary and illuminated by one beam.}\label{sec:_RB1}
\begin{figure}
	\includegraphics[width=15cm]{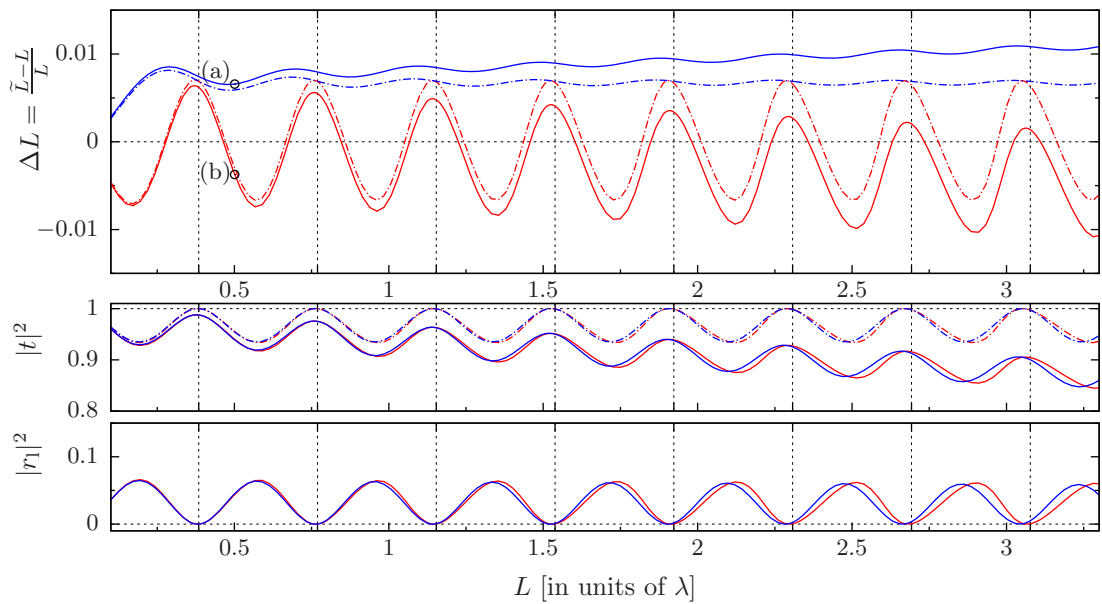}
	\caption{Relative length change, transmission $\abs{t}^2$ and left reflection $\abs{r_\mathrm{l}}^2$ for a dielectric slab fixed at the left edge and illuminated by only one beam incident from the left (blue) or right (red curves) hand side, respectively. The initial refractive index is chosen as $n=1.3+0.0025 \rmi$, the dash-dotted lines show the results for the non-absorptive case $n=1.3$ (in the third plot for $\abs{r_\mathrm{l}}^2$ the additional lines were omitted to avoid ambiguity). The beams are of intensity $I_{\mathrm{l},\mathrm{r}}=0.1 \mathcal{E} c$, the circles indicate the values used in figure~\ref{fig:_Bsp_RB1}. The grid lines mark the locations of the minima [maxima] of the reflectivity [transitivity] in the homogeneous case at $L= m \lambda/(2 \re[n])$, $m\in \mathbb{N}$.}
	\label{fig:_dL_RB1}
\end{figure}
Figure~\ref{fig:_dL_RB1} shows the relative length change in the case where the left edge of a dielectric slab is fixed by some external mechanism. Here we observe a striking difference whether the object is illuminated from the left (blue) or right hand side (red curves). In the first case we find only stretching behaviour with minor oscillations of the relative length change $\Delta L$. However, if the object is illuminated from the right hand side (\ie the beam is incident on the free surface) we again find both compression and elongation, depending on the initial length $L$.\par
Furthermore we observe a stronger impact of absorption than for the previous case with two counterpropagating beams. Now we see that if $\im[n]\neq 0$, then $\Delta L$ increases (light incident from the left, radiation pressure pushing the object to the right) or decreases (light incident from the right) for larger initial lengths $L$. As we see from the dash-dotted lines, the length change continues to oscillate around a constant value also for large values of $L$, if $\im[n]= 0$.\par
For the reflectivity [transitivity], the self consistent strain again results in a shift of the minimal [maximal] values to the left of $L= m \lambda/(2 \re[n])$, $m\in\mathbb{N}$. Note that for the given, asymmetric setup $r_\mathrm{l} \neq r_\mathrm{r}$, but the difference in the total reflectivity $\abs{r_\mathrm{l}}^2-\abs{r_\mathrm{r}}^2$ remains very low for the given parameters.\par
\begin{figure}
	\includegraphics[width=15cm]{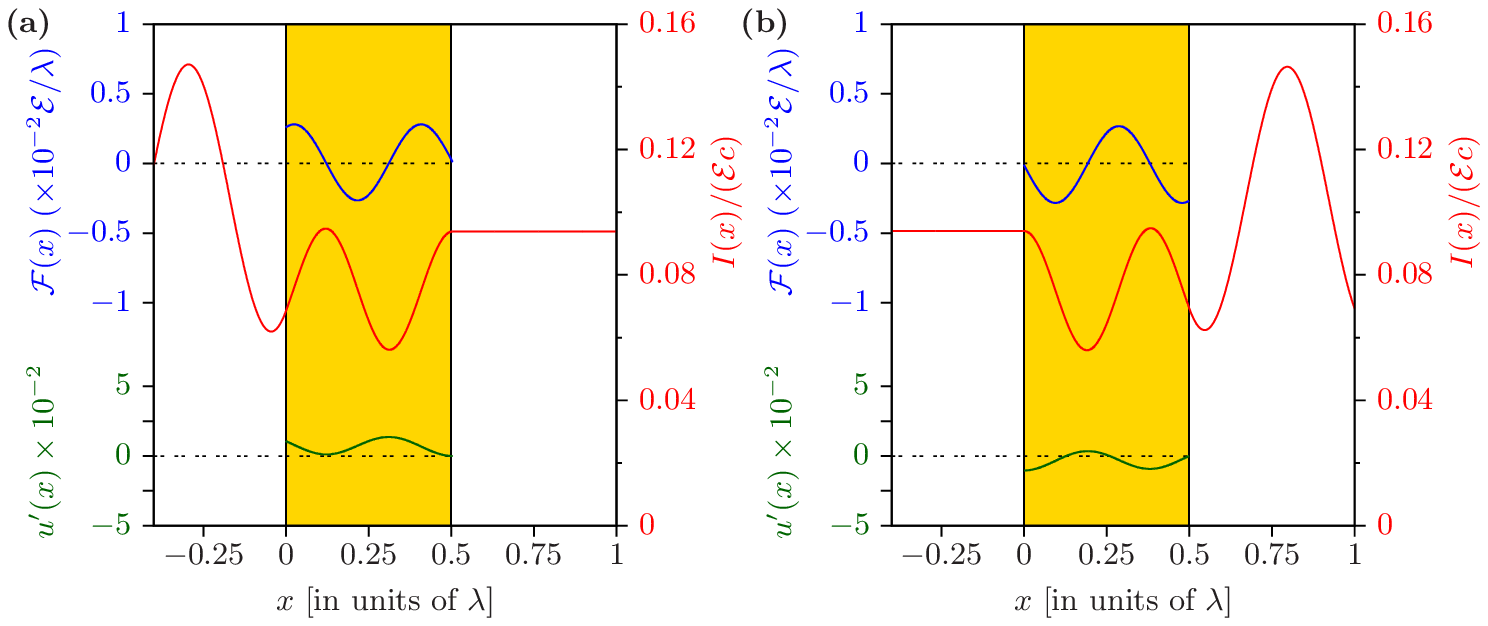}
	\caption{Two dielectric objects fixed at the left edge and illuminated from the left~(a) or from the right hand side only~(b), with $I_{\mathrm{l},\mathrm{r}}=0.1 \mathcal{E} c$. As predicted from~\eref{eq:_Inten_an_Rand_einseitiger_Einfall}, the intensity (red) is locally maximal and the force densities (blue lines) are (almost) zero at the interface where light exits the slab. For left incidence we observe only positive (\ie stretching) strain $u'$ (green lines) whereas both elongating and compressive deformation is feasible for right illumination.}
	\label{fig:_Bsp_RB1}
\end{figure}
To explain the different results for left and right incident beams, we take a closer look at the interface where a single beam exits the slab, \eg at hand of figure~\ref{fig:_Bsp_RB1}. Let us assume a beam entering from the left and $I_\mathrm{r}=0$. Then the intensity is not only constant on the right of the object, but Fresnel's formulae also tell us (for a homogeneous object)
\begin{equation}\label{eq:_Inten_an_Rand_einseitiger_Einfall}
	\tder{x}{}I(L) = 0 \quad \mathrm{and} \quad \tder{x}{2}I(L)=k^2 I(L)(1-2 \re[n]^2+2\im[n]^2) \leq 0,
\end{equation}
and hence the intensity has a local maximum at the edge where the beam exits the slab. For the usually dominant gradient force $\mathcal{F}_\mathrm{gr}(x)\sim \tder{x}{} I(x)$ we thus find $\mathcal{F}_\mathrm{gr}(L) = 0$ and $\mathcal{F}_\mathrm{gr}>0$ left of the surface. Considering the steady state equation~\eref{eq:_Gleichgewicht_Licht_Elast} and omitting the scattering force shows that
\begin{equation}\label{eq:_Dipolkraft_schiebt_zum_Ausgang}
	\mathcal{F}(L)+\mathcal{E} u''(L) \simeq \mathcal{F}_\mathrm{gr}(L)+\mathcal{E} u''(L)=0
\end{equation}
and hence the strain $u'$ has a local minimum at the right edge $x=L$. If light enters only from the right hand side we find analogous behaviour for the left edge $x=0$. Since minimal strain corresponds to a maximum in local material density~\eref{eq:_strain_and_density}, we conclude that the gradient force tends to accumulate material at the surface where a single light beam exits the object. A similar statement holds for the aforementioned case with two counterpropagating beams oscillating in orthogonal polarisations, but then one has to add up the forces generated by the two beams.\par
In total, the constraints on the strain as derived from~\eref{eq:_Inten_an_Rand_einseitiger_Einfall},~\eref{eq:_Dipolkraft_schiebt_zum_Ausgang} and chapter~\ref{sec:_Gleichgewicht_Licht_Elast} are found as
\begin{eqnarray}
	\fl \mathrm{left\,incidence\,only,\,}I_\mathrm{r}=0:\quad &u'(0)=\mathcal{F}_\mathrm{tot}/\mathcal{E} \geq 0; \qquad u'(L) = 0,\,\mathrm{is\,a\,minimum},\\
	\fl \mathrm{right\,incidence\,only,\,}I_\mathrm{l}=0:\quad &u'(0)=\mathcal{F}_\mathrm{tot}/\mathcal{E} \leq 0,\,\mathrm{is\,a\,minimum}; \qquad u'(L)=0.
\end{eqnarray}
So if $I_\mathrm{l}=0$, the strain is fixed at a minimum with negative value on the left edge, at zero on the right edge, and oscillates proportional to the intensity in between. So in total we find both negative (compressive) and positive (expanding) deformation, depending on the length and refractive index of the object. For $I_\mathrm{r}=0$ we see a positive strain at the left edge and a minimum with $u'(L)=0$ at the right boundary. Hence the deformations oscillate between zero and some positive value and always lead to a total stretching behaviour.
\subsection{Identifying length changes by probing the reflectivity}
\begin{figure}
	\includegraphics[width=15cm]{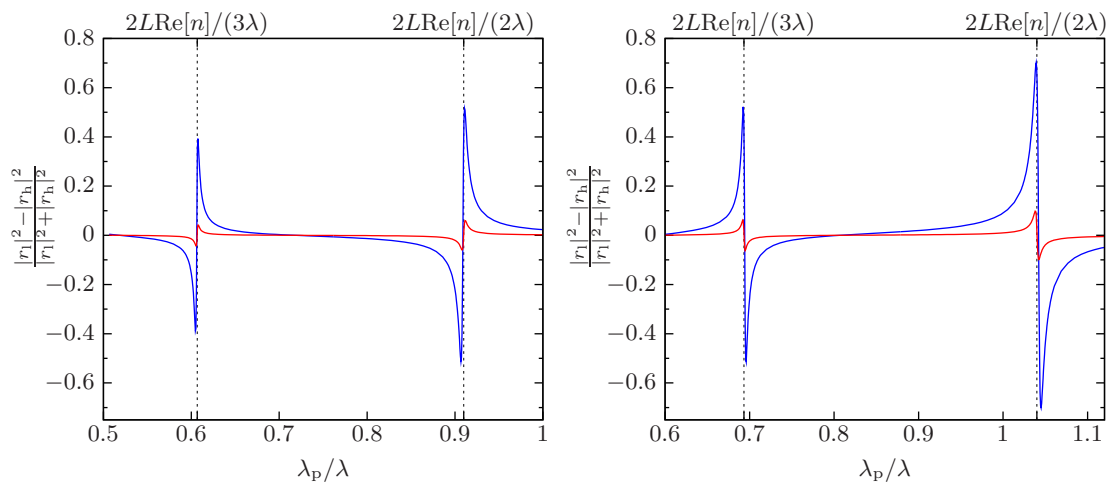}
	\caption{Relative change in the reflectivity of a strained object for different probe laser wavelengths $\lambda_\mathrm{p}$, for slabs trapped in a standing wave pattern with $I_\mathrm{l}=I_\mathrm{r}=0.0005 \mathcal{E} c$ (red) and $I_\mathrm{l}=I_\mathrm{r}=0.005 \mathcal{E} c$ (blue curves), $n=1.3+0.0025  \rmi$. In the left frame we used an initial length $L= 0.7 \lambda$ and obtained compressive behaviour with $\Delta L\simeq-0.2\times 10^{-3}$ (red) and $\Delta L\simeq-2.0\times 10^{-3}$ (blue). The results on the right indicate stretching with $\Delta L\simeq 0.3\times 10^{-3}$ (red) and $\Delta L\simeq 3.1\times 10^{-3}$ (blue) for $L=0.8 \lambda$.}
	\label{fig:_drrt}
\end{figure}
As one can see from figures~\ref{fig:_dL_RB0} and~\ref{fig:_dL_RB1}, the relative length change $\Delta L$ obtained for a given ratio of total intensity to Young's modulus can be imperceptibly small, especially if the original length is not chosen in an optimal relation to the vacuum wavelength of the trapping beam, $\lambda$. But one possibility to detect minor stretching or squeezing for arbitrary initial lengths can be found in the use of a second, weaker laser probing the change in the reflectivity of the medium: Assuming a non-dispersive medium, a probe laser with a vacuum wavelength matching the Fabry-P\'{e}rot condition
\begin{equation}\label{eq:_Bedingung_ReflZero}
	\lambda_\mathrm{p}\simeq \frac{2 L\re n}{m}, \quad m \in \mathbb{N},
\end{equation}
will travel through a slab of length $L$ without being reflected, \ie $\abs{r_\mathrm{h}(\lambda_\mathrm{p})}^2\simeq 0$. Turning on a powerful laser will deform the object and hence also change the reflectivity for the weak probe beam. Figure~\ref{fig:_drrt} shows the relative change of reflectivity
\begin{equation}
	\delta R = \frac{\abs{r_\mathrm{l}}^2-\abs{r_\mathrm{h}}^2}{\abs{r_\mathrm{l}}^2+\abs{r_\mathrm{h}}^2}
\end{equation}
between the unperturbed and the strained medium for different wavelengths of the probe laser beam. We can see that as $\lambda_\mathrm{p}$ crosses values defined in~\eref{eq:_Bedingung_ReflZero}, $\delta R$ changes from negative to positive values if the object is compressed (\ie $\Delta L<0$) and vice versa if $\Delta L>0$.
%
\section{Estimating the deformation by computing the photon momentum transfer on a surface}\label{sec:_photon_momentum}
There exist numerous experimental and theoretical papers reporting optical stretching of deformable objects, such as biological cells~\cite{guck2001optical,rancourt2010dynamic}, or light induced outward bending of liquid-gas surfaces~\cite{ashkin1973radiation,casner2001giant}. In the mentioned publications, the light-induced deformation is estimated by considering an effective photon momentum change at the transition from one medium to another. In this context, the optical forces emerge as surface forces, acting on the interface between two regions of different refractive index. Since the considered materials are incompressible, the refractive index in each region remains constant.\par
Let us try here a similar approach to estimate the deformation of an elastic object and put our results in context with these earlier works. But note that the very different physical properties of a linear elastic medium as compared to incompressible water, plane waves instead of Gaussian beams and a wave description instead of geometric optics, do not allow a straightforward comparison of the results. But nevertheless we can investigate whether our findings based on a volumetric description of optical forces are compatible with a concept of surface forces due to photon momentum exchange.\par
\begin{figure}
	\includegraphics[width=15cm]{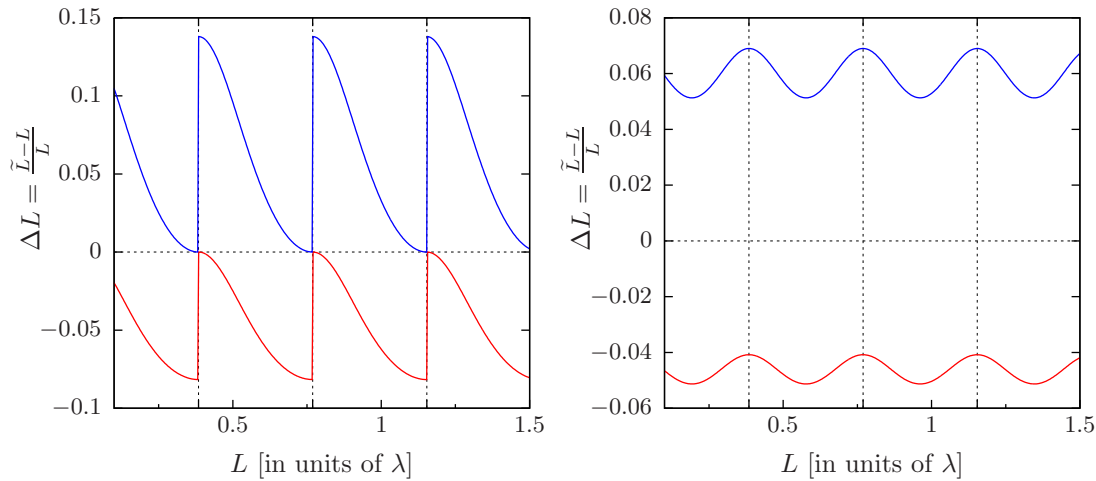}
	\caption{Expected deformation due to the change in photon momentum at the interfaces between vacuum and a homogeneous dielectric of refractive index $n=1.3$, as obtained from~\eref{eq:_deform_force_photonmomentum}. For the blue lines we used Minkowski's photon momentum $p=\hbar k n$, the red curves are computed using Abraham's $p=\hbar k/n$. The left figure shows the situation where a homogeneous object is trapped by a sanding wave with $I_\mathrm{l}=I_\mathrm{r}=0.05 \mathcal{E} c$. As in figure~\ref{fig:_dL_RB0}, the deformation changes abruptly, if the object switches from a low-field to a high-field seeking behaviour. The right hand figure depicts the relative length change for an object illuminated from only one side with $I_\mathrm{l}=0.1\mathcal{E} c$, $I_\mathrm{r}=0$. In both situations we find that Minkowski's momentum only leads to stretching, whereas the naive adaption of Abraham's photon momentum would solely result in compression.}
	\label{fig:_def_photon_momentum}
\end{figure}

Following the line of~\cite{guck2001optical,rancourt2010dynamic,ashkin1973radiation} one estimates the time averaged force per area on an interface separating two regions with different indices of refraction $n_1\neq n_2 \in \mathbb{R}$ and fields $E(x) =A \exp(\rmi n_1 k x) + B \exp(-\rmi n_1 k x)$, $x\leq 0$, and $E(x) =C \exp(\rmi n_2 k x)$, $x> 0$, as
\begin{equation}\label{eq:_force_on_interface}
	F_{1,2} =\frac{\Delta p}{\Delta t} = \frac{I_\mathrm{inc}\big(p_1 (1+R) - p_2 T\big)}{\hbar k c}.
\end{equation}
Here $I_\mathrm{inc}=n_1 c \varepsilon_0\abs{A}^2/2$ is the total energy-flux density entering the system, $R=\abs{B}^2/\abs{A}^2$ and $T=n_2 \abs{C}^2/(n_1 \abs{A}^2)$ is the reflected and transmitted fraction the energy-flux and $p_i=\hbar k n_i$ describes the momentum of a single photon in a medium with index $n_i$, $i=1,2$. As in~\cite{guck2001optical,rancourt2010dynamic,ashkin1973radiation} we here used Minkowski's version of the momentum of light in dielectric media. Out of curiosity about simple but puzzling arguments on stretching or compression of media in connection with the Abraham-Minkowski controversy~\cite{mansuripur2010resolution,ashkin1973radiation,pfeifer2007colloquium,barnett2010enigma} we also included results computed by naively inserting Abraham's result for the photon momentum, $p_i=\hbar k/n_i$, in figure~\ref{fig:_def_photon_momentum}.
One must note, however, that Abraham's stress tensor would also require a material tensor component. A thorough calculation should always give the same results, independent of the used version of stress tensor~\cite{pfeifer2007colloquium,barnett2010enigma}.\par
Neglecting internal reflections, the force on an extended object with index $n_2$ embedded in a medium $n_1$ and subjected to a single beam is estimated to give $F_\mathrm{tot}=F_{1,2}+T F_{2,1}$. The deformation of such an object is then simply the difference in the two forces on the surfaces, reading $F_\mathrm{def}=T F_{2,1}-F_{1,2}$. Assuming a linear elastic medium with Young's modulus $\mathcal{E}$, the relative length-change can be estimated by $\Delta L =F_\mathrm{def}/\mathcal{E}$. For the values used in the previous examples $n_1=1$, $n_2=1.3$, $I_0=0.1 \mathcal{E} c$, we obtain $\Delta L\simeq 0.059$ when using Minkowski's momentum and $\Delta L \simeq -0.046$ for Abraham. These deformations have about the same order of magnitude as our full self-consistent computations, depicted \eg in figure~\ref{fig:_dL_RB1}, but do not depend on the length of the object.\par
Formally one can refine these calculations from~\eref{eq:_force_on_interface} and include also light incident from the right such that $E(x) =C \exp(\rmi n_2 k x)+ D \exp(-\rmi n_2 k x)$, $x> 0$ to obtain
\begin{equation}\label{eq:_force_on_interface_better}
	F_{1,2} = \frac{\varepsilon_0 n_1 p_1}{2 \hbar k}\big(\abs{A}^2+\abs{B}^2\big) - \frac{\varepsilon_0 n_2 p_2}{2 \hbar k}\big(\abs{C}^2+\abs{D}^2\big).
\end{equation}
For $D=0$ this reduces to~\eref{eq:_force_on_interface} and for $n_1=n_2=1$ and $p_1=p_2=\hbar k$ we recover the force derived previously with the Maxwell stress tensor~\eref{eq:_Kraft_Fj}. This allows to formulate a generic wave optics extension for the deformation estimated above in the scope of geometric optics, now including also size dependent reflection and transmission. The total deformation pressure on an object with length $L$ and homogeneous refractive index $n \in \mathbb{R}$ surrounded by vacuum then reads
\begin{equation}\label{eq:_deform_force_photonmomentum}
	F_\mathrm{def}=\frac{\varepsilon_0 n p_n}{\hbar k}\big(\abs{G}^2+\abs{H}^2\big) - \frac{\varepsilon_0}{2}\big(\abs{A}^2+\abs{B}^2-\abs{C}^2-\abs{D}^2\big)
\end{equation}
where the amplitudes inside the medium are computed using Fresnel's relations $G=\big((n+1)A+(n-1)B\big)/(2n)$, $H=\big((n-1)A+(n+1)B\big)/(2n)$, the amplitudes outside are connected by the homogeneous reflection and transmission coefficients~\eref{eq:_r_und_t_homogen}, $B=r_\mathrm{h} A+t_\mathrm{h} D$, $C=t_\mathrm{h} A+r_\mathrm{h} D$, and the incoming $A$ and $D$ are given in~\eref{eq:_A0_DL_tilde}. As expected, a similar calculation for the total force $F_\mathrm{tot}$ gives the same result as we obtained previously in~\eref{eq:_Kraft_Med_1}.\par
The resulting relative length change $\Delta L =F_\mathrm{def}/\mathcal{E}$ is presented in figure~\ref{fig:_def_photon_momentum}. There we find that the estimations using optical surface forces even qualitatively differ from the results we obtained with the present description using the full, volumetric optical forces, \cf figures~\ref{fig:_dL_RB0} or~\ref{fig:_dL_RB1}, even if the force on the surface is adapted to include interference due to internal reflections.\par
An intuitive example is the object of length $L=\lambda/(2 n)$ where $r_\mathrm{h}=0$ and $t_\mathrm{h}=1$. In a standing wave trap with $I_l=I_r$ this object is then trapped at $\xi_0=x_0 + (\phi_\mathrm{l} - \phi_\mathrm{r})/(2 k)=0$, \cf chapter~\ref{sec:_Fallenposition} or figure~\ref{fig:_trap_stiffness}. Using Minkowski's $p_n=\hbar k n$ we then find that the force due to photon momentum transfer vanishes at each surface and hence $\Delta L=0$ in figure~\ref{fig:_def_photon_momentum}.\par
But from the examples in figure~\ref{fig:_Bsp_RB0_Int1} we deduce that the intensity at the \emph{surface} is zero, yet the object will contract due to the intensity maximum at its central position. This is because the dipole force pulls each volume element towards the next local maximum of intensity, regardless of whether this volume element is located at the surface or in the bulk of the medium.\par
In fact, none of our calculations or simulations showed any distinctive effects suggesting a surface force at the boundaries of a dielectric~\cite{sonnleitner2011optical}. This is  supported by computations on large but finite stacks of polarizable slices where the forces on the first or last slice qualitatively do not differ from those on the second or next to last, respectively. We therefore conclude that optical forces have to be treated as real volumetric forces~\cite{mansuripur2010resolution,rinaldi2002body} and that a description using the change of photon momentum at the surface of a medium is inappropriate, regardless of using Abraham's or Minkowski's momentum.
%
\section{Conclusions}
Using an implicit calculation of optical fields and forces allows to self-consistently determine the stationary local deformations of an elastic object, where the local stress balances the local light forces by elastic back action. These solutions show a surprisingly variable and nonlinear dependence on the chosen parameters. Generally we see a length and illumination dependent, spatially quasiperiodic strain pattern, which can lead to length stretching as well as compression. As expected, standing wave configurations yield the strongest forces and effective length changes with a clear resonant structure for special ratios of initial object length $L$ and trap beam wavelength $\lambda$. In the standing wave setup, variations in the trap wavelength $\lambda$ lead to discrete jumps of the stable trapping positions. At $L= m \lambda/(2 \re[n])$, $m\in \mathbb{N}$, the particle switches from a position centred around an intensity maximum to one around a field node, which is associated with changes from compression to elongation of the object. Interestingly, in particular close to these instability points, this generally leads to an increase in trap stiffness. We expect that this indicates possible bistability between high and low field seeking behaviour for certain lengths very close to $L= m \lambda/(2 \re[n])$.\par
Although the calculations presented here were performed in the scope of elastic media, we believe that the model can be extended to deformable but incompressible media like water or even dilute gases. Here in particular stability thresholds for the homogeneous solutions should prove physically very interesting, as they could lead to stationary flows, periodic density oscillations or light induced density pattern formation and particle ordering in a gas.\par
Here we limited our considerations to the case, where a steady state solution can be found. As for other nonlinear dynamical effects~\cite{asboth2008optomechanical}, there are of course regions in parameter space, were no stationary solutions exist and we find self sustained oscillations or even disintegration of the material. Indications of this behaviour appear \eg in a non converging iteration procedure. At this point we leave this to future work.
\ack
We thankfully acknowledge support via ERC Advanced Grant (catchIT, 247024) and the Austrian Science fund FWF grant S4013.
%
\appendix
\section{General features of transfer matrices}\label{sec:_TrafoMatrizen}
In equation~\eref{eq:_TrafoMatrix_beamsplitter} we already used the concept of a transfer matrix to couple the plane wave amplitudes left and right of a beam splitter. Let us generally define the set of transfer matrices as
\begin{equation}\label{eq:_Def_Transfermatrix} \fl
	\mathcal{T}:= \bigg\{ \mathrm{T} \in \mathbb{C}^{2 \times 2} \; \Big \vert \; \exists\ r_1,r_2,t\in\mathbb{C}:
 \mathrm{T} = \frac{1}{t}
	\pmatrix{
		t^2 - r_1 r_2	&	r_2	\cr
		- r_1			&	1 }
 \equiv \mathrm{T}(r_1,r_2,t)\bigg \}.
\end{equation}
One can easily show that $\mathrm{T}_1 \cdot \mathrm{T}_2 \in \mathcal{T}$ for all $\mathrm{T}_1, \mathrm{T}_2 \in \mathcal{T}$ and $\cdot$ here denoting the usually omitted matrix multiplication. Since also
\begin{equation}
	[\mathrm{T}(r_1,r_2,t)]^{-1}=\frac{1}{t} \pmatrix{
			1		&	r_1				\cr
			- r_2	&	t^2 - r_1 r_2	} 
	\in \mathcal{T}
\end{equation}
we conclude that $(\mathcal{T},\cdot)$ is a group.\par
To motivate definition~\eref{eq:_Def_Transfermatrix}, let us consider two plane waves $E_\mathrm{l}(x)=A\exp(\rmi k (x))+B\exp(-\rmi k (x))$, $x\leq 0$, and $E_\mathrm{r}(x)=C\exp(\rmi k (x-L))+D\exp(-\rmi k (x-L))$, $x\geq L$, left and right of a dielectric with a homogeneous refractive index $n$. Hence their amplitudes are coupled as
\begin{equation}
	\pmatrix{ C \cr D } = \mathrm{S}_{n,1}\cdot\mathrm{P}_{n L}\cdot\mathrm{S}_{1,n} \pmatrix{ A \cr B }.
\end{equation}
where $\mathrm{S}_{n_1,n_2}$ couples the amplitudes at the intersection from a region with refractive index $n_1$ to a region with index $n_2$ and $\mathrm{P}_{d}$ denotes the propagation matrix over a distance $d$, \ie
\begin{equation}
	\mathrm{S}_{n_1,n_2}= \frac{1}{2 n_2}
		\pmatrix{
			n_2+n_1	&	n_2-n_1	\cr
			n_2-n_1	&	n_2+n_1	}
	\quad \mathrm{and}\quad
	\mathrm{P}_{d}=\pmatrix{
			\rme^{\rmi k d}	&	0	\cr
			0	&	\rme^{-\rmi k d}}.
\end{equation}
One can easily show that $\mathrm{S}_{n,1}\cdot\mathrm{P}_{n L}\cdot\mathrm{S}_{1,n} = \mathrm{T}(r_\mathrm{h},r_\mathrm{h},t_\mathrm{h})$, with $r_\mathrm{h}$ and $t_\mathrm{h}$ as given in~\eref{eq:_r_und_t_homogen}. But the attempt to construct a total transfer matrix for stacked media such as $\mathrm{S}_{n_K,1}\cdot\mathrm{P}_{n_K d_K}\cdot\mathrm{S}_{n_{K-1},n_K}\cdots\mathrm{S}_{n_2,n_1}\cdot\mathrm{P}_{n_1 d_1}\cdot\mathrm{S}_{1,n_1}$ with arbitrary $n_i$, $d_i$, $i=1,\ldots, K$, will show that one reflection coefficient is not enough. However, such a system can be described by a more general $T(r_1,r_2,t) \in \mathcal{T}$ such that $B=r_1 A+ t D$ and $C=r_2 D+t A$.
%
\section{Analytical approximations for electric fields and forces for small deformations}\label{sec:_ApproxField_Force}
In section~\ref{sec:_inhom_Felder} and in an earlier work~\cite{sonnleitner2011optical} we showed that for equally spaced slices the amplitudes of the electric fields are related as $(A_{j+1}, B_{j+1})^T = \mathrm{T_h}^j (A_1, B_1)^T$, with $\mathrm{T_h}:=\mathrm{P}_{d_0} \mathrm{M_{BS}}$. Choosing the coupling $\zeta$ as in~\eref{eq:_zeta}, the eigenvalues of $\mathrm{T_h}$ read $\exp(\pm \rmi n k d_0)$. This leads to
\begin{equation}\label{eq:_Trafomatrix_Grenzwert}
	\lim_{N\rightarrow \infty} \mathrm{T_h}^{j-1} =: \mathrm{T}(x) = \frac{1}{2 n}
		\pmatrix{
			f(-x)+g(-x)	&	f(x)-g(x)	\cr
			f(-x)-g(-x)	&	f(x)+g(x)	}	
\end{equation}
where $x=\lim_{N\rightarrow\infty} L (j-1)/(N-1)$ and 
\begin{equation} \label{eq:_Def_f_und_g} \eqalign{
	f(x) = n \cos(n k x)-\rmi \sin(n k x),	\\
	g(x) = n \cos(n k x)-\rmi n^2 \sin(n k x).
}\end{equation}
Allowing small local density variations $\widetilde{d}_j-d_0 =\Delta_j$ we may use $\mathrm{P}_{d_j} \simeq \mathrm{P}_{d_0}+ \rmi k \Delta_j \sigma_z \mathrm{P}_{d_0}$ to expand the relation between the field amplitudes from~\eref{eq:_Amplitudenrelation_BS_allgemein} as
\begin{equation} \label{eq:_Trafomatrix_inhom_naeherung_diskret}
	\pmatrix{ A_{j+1} \cr B_{j+1} }
	\simeq
	\Big[\mathrm{T_h}^{j} + 
		\rmi k \sum_{m=1}^j \Delta_m \mathrm{T_h}^{j-m}\sigma_z \mathrm{T_h}^{m} \Big]
	\pmatrix{ A_{1} \cr B_{1} },
\end{equation}
where $\sigma_z=\diag(1,-1)$. In the limit of infinitely many slices within a finite length $L$, the sum above can be rewritten to an integral and we obtain
\begin{equation}\label{eq:_Trafomatrix_inhom_naeherung_kont}
	\pmatrix{ A(x) \cr B(x) }
	 \simeq 
		\Big[\mathrm{T}(x) + \rmi k \int_0^x u'(y) \mathrm{T}(x-y) \sigma_z \mathrm{T}(y) \rmd y \Big]
		\pmatrix{ A(0) \cr B(0) }.
\end{equation}
Since $\lim_{N\rightarrow \infty}\mathrm{M_{BS}}=\mathrm{id}_2$, the total reflection and transmission coefficients for the displaced object can be read off and expanded in linear order of $u^\prime$ from equations~\eref{eq:_Einfuehrung_rrt} and~\eref{eq:_Trafomatrix_inhom_naeherung_kont}
\begin{equation}\label{eq:_r_und_t_inhomogen}\eqalign{
	t 	\simeq t_\mathrm{h} + \case{2 \rmi k}{(f(L)+g(L))^2} \int_{0}^L u^\prime(y) \big(f(y)f(L-y)+g(y)g(L-y)\big) \rmd y,	\\
	r_\mathrm{l} \simeq r_\mathrm{h} + \case{2 \rmi k}{(f(L)+g(L))^2} \int_{0}^L u^\prime(y) \big(f(L-y)^2-g(L-y)^2\big) \rmd y, \\
	r_\mathrm{r} \simeq r_\mathrm{h} + \case{2 \rmi k}{(f(L)+g(L))^2} \int_{0}^L u^\prime(y) \big(f(y)^2-g(y)^2\big) \rmd y.
}\end{equation}
Here $r_\mathrm{h}$ and $t_\mathrm{h}$ denote the reflection and transmission amplitudes of a homogeneous medium with length $L$ and refractive index $n$, as given in~\eref{eq:_r_und_t_homogen}. As mentioned in section~\ref{sec:_rrt} we find that symmetric local strain, \ie $u'(x)=u'(L-x)$, $x\in[0,L]$, results in $r_\mathrm{l}\simeq r_\mathrm{r}$ and antisymmetric strain gives $t\simeq t_\mathrm{h}$ and $r_\mathrm{l}-r_\mathrm{h} \simeq r_\mathrm{h}-r_\mathrm{r}$.\par
Using the reflection and transmission amplitudes from~\eref{eq:_r_und_t_inhomogen} to replace $B(0)=r_\mathrm{l} A(0)+t D(L)$ in~\eref{eq:_Trafomatrix_inhom_naeherung_kont} leads to analytical approximations for the local field amplitudes inside a deformed medium
\begin{equation}\label{eq:_Felder_kurz}\eqalign{
	A(x) \simeq a_\mathrm{h}(x) + \int_0^x u'(y) a_\mathrm{x}(x,y) \rmd y + \int_0^L u'(y) a_\mathrm{L}(x,y) \rmd y,	\\
	B(x) \simeq b_\mathrm{h}(x) + \int_0^x u'(y) b_\mathrm{x}(x,y) \rmd y + \int_0^L u'(y) b_\mathrm{L}(x,y) \rmd y,
}\end{equation}
with the individual terms reading
\begin{equation}\label{eq:_Felder_Einzelterme}\eqalign{
\fl	a_\mathrm{h}(x) = \frac{A_0 \big(f(L-x)+g(L-x)\big)+D_L \big(f(x)-g(x)\big)}{f(L)+g(L)}, \\
\fl	b_\mathrm{h}(x) = \frac{A_0 \big(f(L-x)-g(L-x)\big)+D_L \big(f(x)+g(x)\big)}{f(L)+g(L)}, \\
\fl	a_\mathrm{x}(x,y) = \frac{\rmi k}{n \big(f(L)+g(L)\big)}
		\Big(A_0\big(f(L-y)f(y-x)+g(L-y)g(y-x)\big) + \\
		 \qquad\qquad\qquad\qquad + D_L\big(f(y)f(y-x)-g(y)g(y-x)\big)\Big),\\
\fl	b_\mathrm{x}(x,y) = \frac{- \rmi k}{n \big(f(L)+g(L)\big)}
		\Big( A_0\big(f(L-y)f(x-y)-g(L-y)g(x-y)\big) + \\
		 \qquad\qquad\qquad\qquad + D_L\big(f(y)f(x-y)+g(y)g(x-y)\big)\Big), \\
\fl	a_\mathrm{L}(x,y) = \frac{\rmi k \big(f(x)-g(x)\big)}{n \big(f(L)+g(L)\big)^2} 
		\Big( A_0\big(f(L-y)^2-g(L-y)^2\big) + \\
		 \qquad\qquad\qquad\qquad + D_L \big(f(y)f(L-y)+g(y)g(L-y)\big)\Big), \\
\fl	b_\mathrm{L}(x,y) = \frac{\rmi k \big(f(x)+g(x)\big)}{n \big(f(L)+g(L)\big)^2} 
		\Big( A_0\big(f(L-y)^2-g(L-y)^2\big) + \\
		 \qquad\qquad\qquad\qquad + D_L \big(f(y)f(L-y)+g(y)g(L-y)\big)\Big).
}\end{equation}
The functions $f$ and $g$ are given in~\eref{eq:_Def_f_und_g}, the amplitudes at the boundary $A_0\equiv A(0)$ and $D_L\equiv D(L)$ are determined by the incoming intensities and the displacement and can be read off from~\eref{eq:_A0_DL_tilde}.\par
To obtain an approximation for the forces we use~\eref{eq:_Kraft_Fj_anders}, take the limit $\mathcal{F}(x):= \lim_{N\rightarrow \infty} N F_j/L$ with $\lim_{N \rightarrow \infty}N\zeta/L= k(n^2-1)/2$ and insert the amplitudes from~\eref{eq:_Felder_kurz} to find
\begin{eqnarray}
	\mathcal{F}(x) &= \case{k \varepsilon_0}{2} \im\big\{(n^2-1)\big(A(x)+B(x)\big) 
											\big(A(x)-B(x)\big)^\ast \big\} \nonumber	\\
	\mathcal{F}(x) &\simeq \mathcal{F}_\mathrm{h}(x) + \int_0^x u'(y)\mathcal{F}_\mathrm{x}(x,y) \rmd y + \int_0^L u'(y)\mathcal{F}_\mathrm{L}(x,y) \rmd y,
\label{eq:_Kraftdichte_anayltisch_inhom}
\end{eqnarray}
where
\begin{equation}\label{eq:_Kraftdichte_Einzelterme}\eqalign{
\fl	\mathcal{F}_\mathrm{h}(x) =
			\case{2 k \varepsilon_0}{\abs{(f(L)+g(L)}^2} \im \big\{ (n^2-1)
 					(A_0 f(L-x) + D_L f(x))(A_0 g(L-x) - D_L g(x))^\ast \big\} \\
\fl	\mathcal{F}_\mathrm{x}(x) =
			\case{k \varepsilon_0}{2}\im \Big\{ (n^2-1) \Big[(a_\mathrm{h}(x)-b_\mathrm{h}(x))^\ast 
					(a_\mathrm{x}(x,y)+b_\mathrm{x}(x,y)) + \\
\fl			\hspace{5cm} (a_\mathrm{h}(x)+b_\mathrm{h}(x)) 
					(a_\mathrm{x}(x,y)-b_\mathrm{x}(x,y))^\ast \Big]\Big\} \\
\fl	\mathcal{F}_\mathrm{L}(x) =
			 \case{k \varepsilon_0}{2}\im \Big\{ (n^2-1) \Big[(a_\mathrm{h}(x)-b_\mathrm{h}(x))^\ast 
					(a_\mathrm{L}(x,y)+b_\mathrm{L}(x,y)) + \\
\fl			\hspace{5cm} (a_\mathrm{h}(x)+b_\mathrm{h}(x))
					(a_\mathrm{L}(x,y)-b_\mathrm{L}(x,y))^\ast \Big]\Big\}
}
\end{equation}
with the terms $a_\mathrm{X}$ and $b_\mathrm{X}$ as given above in~\eref{eq:_Felder_Einzelterme}.\par
Just like the field amplitudes in~\eref{eq:_Felder_kurz} and the reflection and transmission coefficients in~\eref{eq:_r_und_t_inhomogen}, the above result is only an approximation for the case of small deformation $u'$. That is, we neglect products of $\Delta_l \Delta_m$ in~\eref{eq:_Trafomatrix_inhom_naeherung_diskret} and hence also correlations of type $\int\int u'(y_1) u'(y_2) \dots \rmd y_1 \rmd y_2$. But a comparison of the approximated results in the continuous limit with solutions of the wave equation~\eref{eq:_Wellengl_n(x)} or numerical computations for a large but finite number of beam splitters confirmed that this first order expansion is sufficient for the scope of parameters used in this work.
%
%
\section*{References}

\end{document}